\documentclass[pre,aps,onecolumn,showpacs,floatfix,10pt,superscriptaddress]{revtex4-2}

\usepackage{graphicx}
\usepackage{amsmath}
\usepackage{mathtools}
\usepackage{dcolumn}
\usepackage{bm}
\usepackage{subfigure}
\usepackage{epsfig}
\usepackage{afterpage}
\usepackage[colorinlistoftodos]{todonotes}
\usepackage[colorlinks,citecolor=red,urlcolor=blue,bookmarks=false,hypertexnames=true]{hyperref}

\newcommand{\ra}{\rangle}
\newcommand{\la}{\langle}
\newcommand{\bea}{\begin{eqnarray}}
\newcommand{\eea}{\end{eqnarray}}
\newcommand{\beq}{\begin{equation}}
\newcommand{\eeq}{\end{equation}}
\newcommand{\be}{\begin{equation}}
\newcommand{\ee}{\end{equation}}
\newcommand{\beqa}{\begin{eqnarray}}
\newcommand{\eeqa}{\end{eqnarray}}

\newcommand{\lx}{\lambda}

\definecolor{dgreen}{rgb}{0,0.7,0}

\begin{document}
	
\title{Mass fluctuations in Random Average Transfer Process in open set-up}
	\author{Rahul Dandekar}
	\affiliation{Institut de Physique Theorique, CEA, CNRS, Universite Paris–Saclay, F–91191 Gif-sur-Yvette cedex, France}
	\author{Anupam Kundu}
	\affiliation{International Centre for Theoretical Sciences, Bangalore, India}
	
	\date{\today}
	
	\begin{abstract}
		We define a new mass transport model on a one-dimensional lattice of size $N$ with continuous masses at each site. The lattice is connected to mass reservoirs of different `chemical potentials' at the two ends. The mass transfer dynamics in the bulk is equivalent to the dynamics of the gaps between particles in the Random Average Process. In the non-equilibrium steady state, we find that the multi-site arbitrary order cumulants of the masses can be expressed {as an expansion in powers of $1/N$ where at each order} the cumulants have {a} scaling form. We introduce a novel operator approach which allows us to compute these scaling functions at different orders of $1/N$. Moreover, this approach reveals that, to express the scaling functions for higher order cumulants completely one requires all lower order multi-site cumulants. This is in contrast to the Wick's theorem in which all higher order cumulants are  expressed solely in terms of two-site cumulants. We support our results with evidence from Monte-Carlo simulations.
	\end{abstract}
	
	\maketitle
	
	\section{Introduction } 
	\label{sec:intro}
\noindent
Characterization of the steady-state and dynamical properties of systems which are driven out of thermal equilibrium has been a subject of intense theoretical and experimental studies for many years. Unlike their equilibrium counterparts, no general theoretical framework exists within which properties of non-equilibrium systems can be analysed. More precisely, the stationary distribution of a system in a non-equilibrium steady state (NESS) is often not known unlike the equilibrium case where the Gibbs-Boltzmann distribution is known to be correct. To characterise and describe such out-of-equilibrium systems, one often studies the fluctuations and correlations among different degrees of freedom or among some coarse-grained degrees of freedom of the system.

Mass transfer models have played a paradigmatic role in the formulation of theory of non-equilibrium systems. The zero range process \cite{evans2005nonequilibrium}, for example, has played an important role in studying non-equilibrium phase transitions and hydrodynamics. Other examples include chipping models \cite{majumdar2000nonequilibrium,evans2004factorized} 
the misanthrope process \cite{cocozza1985processus,evans2014condensation}, the simple exclusion process \cite{derrida2011microscopic,mallick2011some} and the random average process \cite{ferrari1998fluctuations,krug2000asymmetric,rajesh2001exact,cividini2016exact,cividini2016correlation}. In last decade, a general formulation called Macroscopic Fluctuation Theory \cite{bertini2015macroscopic} has been developed which describes mass transfer models with short range transfer and obeying `gradient condition' \cite{spohn2012large,landim2004hydrodynamic}. In this formulation such systems exhibit (fluctuating) diffusive hydrodynamics of a conserved density field $\rho(x,t)$, characterised solely by the diffusivity $D(\rho)$ and the mobility $\sigma(\rho)$. To obtain the density dependence of the transport coefficients $D(\rho)$ and $\sigma(\rho)$, one usually starts from a microscopic description and identifies the coarse-grained density, corresponding to the microscopic conserved quantity of the dynamics, and the associated current. One calculates $D(\rho)$ and $\sigma(\rho)$ by computing the variance of the local density and its response to a small external field \cite{bertini2015macroscopic,das2017einstein}. The transport coefficients 
calculated in this way only contain information about the first two moments of the mass transfer statistics. It is assumed that higher moments of the mass transfer distribution scale out under coarse-graining and thus do not matter in the large system-size limit.

In this paper, we study a mass transport model which is the dual of the random average process \cite{cividini2016exact}. This model is defined on a one dimensional lattice of size $N$, with each site has positive mass $g_i$ where $i=1,2,...,N$ is the site index. A random fraction $\eta$ of the mass from site $i$ goes with equal probability to the neighbouring sites $i \pm 1$ at unit rate, with the distribution of $\eta$, $R(\eta)$, being arbitrary. In this model, one can investigate the effect of higher moments of the mass-transfer distribution systematically by tuning the distribution $R(\eta)$. Although only the first two moments of $R(\eta)$ appear in the hydrodynamic description of the RAP \cite{ferrari1998fluctuations,kundu2016exact} we show that all higher moments appear in the continuum limit of this process, as one studies cumulants of higher orders in the site masses $g_i$. Thus, we go beyond the hydrodynamic description to provide a microscopic continuum limit of the process.

We study the process in an `open system' setting, where the boundary sites are connected to two reservoirs of different 'chemical potentials'. As a result the system reaches a non-equilibrium steady-state (NESS) with a net current across the system. In this paper we study this NESS by computing various cumulants of the masses $g_i$. In a recent study it has been observed that in the NESS the two point correlations $\la g_i g_j \ra_c = \la g_i g_j \ra  -\la g_i \ra \la g_j \ra$ as well as the single site variance $\la g_i^2 \ra_c =\la g_i^2 \ra - \la g_i \ra^2$ possess scaling behaviour in the large $N$ limit with determinable scaling forms \cite{cividini2016correlation}. We show that the particular structure of the RAP process allows one to extend this scaling behaviour to multi-site cumulants of arbitrary order. Employing a novel operator method we compute these scaling functions in terms of the scaling functions {of} lower order correlation functions.
	
	\section{Definition of the model and summary of the results} 
	\label{sec:model}
	\noindent
	The mass transfer process is defined on a line of $N$ sites, with the masses $g_i,~i \in \{1,2,...,N\}$ at each site being continuous positive variables. The system is connected to mass reservoirs at site $1$ and $N$. The mass from site $1$ $(N)$ can get transferred to the reservoirs on the left (right). Similarly, mass from the reservoir {on} the left (right) can come to site $1 (N)$. The dynamics at any site $i \neq 1, N$ is given by:
	
	\begin{itemize}
		\item[-] In a small time interval $dt$, a random fraction $\eta$ of mass $g_i$ from the site $i$ gets transferred to either of its neighbors $(i-1)$ or $(i+1)$ with equal probability $dt/2$.
		\item[-] The  random variable $\eta \in (0,1)$ is chosen from a given probability distribution $R(\eta)$. 
	\end{itemize}
	
	Mathematically, for a given pair of sites say, $\{i,i+1\}$, in the bulk, the dynamics can be written as 
	\beqa
	\{g_i(t+dt),g_{i+1}(t+dt)\} = 
	\begin{cases}
		\{g_i(t) (1-\eta), g_{i+1} + \eta g_i\} & \text{with prob }  R(\eta) d\eta \frac{dt}{2} \\
		&~~~~~~~~~~~~~~~~~~~~~~~~~~~~\text{for}~~2\leq i \leq N-1, \\
		\{g_i(t),g_{i+1}(t)\} & \text{with prob } 1 - \frac{dt}{2}
	\end{cases}
	\label{mass-dyna}
	\eeqa
	Note that the dynamics in the bulk is mass-conserving and \textit{homogeneous under uniform re-scaling of all the masses}. The sites $1$ and $N$ exchange mass with the reservoirs at site $0$ and $N+1$ respectively. The reservoirs are described as follows. The distribution of the mass $G_L$ at, say the left reservoir at site $0$, is given by a specified distribution $P_L(G_L)$. Similarly the distribution of the mass $G_R$ at the right reservoir at site $N+1$ is given by $P_R(G_R)$. These two distributions are seen as externally controlled and hence remain unchanged with time even though the reservoirs are exchanging mass with the system. The dynamics at the boundary sites $1$ and $N$ is defined as:
	\begin{itemize}
		\item[-] In every time step $dt$, site $1$ transfers a random fraction $\eta$ of its mass to the right to site 2 with probability $dt/2$ or to the left boundary of the system, with probability $dt/2$, where it disappears from the system. Also, in a time $dt$, with probability $dt/2$, a mass $\eta G_L$ can get transferred to site $1$.
		\item[-] Site $N$ behaves similarly, either losing mass $\eta g_N$ to the right boundary with probability $dt/2$ or gaining a mass $\eta G_R$ with probability $dt/2$ from the right boundary. We take the distribution of the mass in the reservoirs to be same as the steady state distribution of masses when there is no drive \cite{cividini2016exact}. 
		
	\end{itemize}
	We call this mass transfer process the random average transfer process (RATP) as this process is closely related to the random average process (RAP). This connection is best illustrated in the ring geometry. The RAP process \cite{ferrari1998fluctuations,krug2000asymmetric} on a ring is defined with $N$ single file particles moving on a 1D ring, which can not overtake each other. In the random-sequential version of the process, in a small time step $dt$, a particle, say the $i^{\rm{th}}$, jumps from its current position $x_i$ either to the left or to the right with equal probability $dt/2$. Whenever the particle jumps in one direction, it jumps by a random fraction of the space available till the next particle in that direction. As a result, they maintain their initial order of sequence and do not overtake each other. Now, corresponding to the $N$ particles in the RAP, consider a periodic one dimensional lattice of $N$ sites performing mass transfer where the mass at the $i^{\rm{th}}$ site is exactly the gap $g_i=x_{i+1}-x_i$ between the $i^{\rm{th}}$ and $(i+1)^{\rm{th}}$ particle in the RAP picture. Thus particles from the RAP picture are mapped to the links between lattice sites in the RATP picture. Now, the jump made by the  $i^{\rm{th}}$ particle towards the $(i+1)^{\rm{th}}$ particle in the RAP picture corresponds to mass transfer from the $i^{\rm{th}}$ site to $(i-1)^{\rm{th}}$ site in the RATP picture.  This establishes the connection between the single-file process and the mass transfer process and also justifies the name RATP.
	
	The state of the RATP system at any time $t$ is given by the masses $\{g_i\}$ at the sites $1 \le i \le N$. Let the probability distribution of this configuration is denoted by $P(\{g_i\},t)$. This distribution evolves according the following master equation 
	\beqa
	\frac{d}{dt} P(\{g_i\}) = &\frac{1}{2 N}& \left[ \sum_{i=1}^{N-1} \int_{0}^{\infty} dg_i' \int_{0}^{\infty} dg_{i+1}' \int_{0}^{1} d\eta R(\eta) P(\{g_i'\}) \right. \nonumber \\ & \times & \{\delta(g_i - g_i' + \eta g_i')\delta(g_{i+1}- g_{i+1}' - \eta g_i) + \delta(g_{i+1} - g_{i+1}' + \eta g_{i+1}')\delta(g_{i}- g_{i}' - \eta g_{i+1})\} \nonumber\\ &+& \int_{0}^{\infty} dg_1' \int_{0}^{\infty} dG_L \int_{0}^{1} d\eta R(\eta) P(\{g_i'\}) P_L(G_L) \{\delta(g_1 - g_1' + \eta g_1') + \delta(g_1 - g_1' - \eta G_L)\} \nonumber\\ &+& \left. \int_{0}^{\infty} dg_N' \int_{0}^{\infty} dG_R \int_{0}^{1} d\eta R(\eta) P(\{g_i'\}) P_R(G_R) \{\delta(g_N - g_N' + \eta g_N') + \delta(g_N - g_N' - \eta G_R)\} \right],
	\label{eq:Pt}
	\eeqa
	where $\{g_i'\}$ is the configuration $\{g_i\}$ with $g_i, g_{i+1}$ replaced with $g_i',g_{i+1}'$, and $P_L(G)$ is a distribution with mean $g_L$ and $P_R(G)$ has mean $g_R$. When $g_L \neq g_R$, the system is driven by the particle reservoirs at the two ends. As a result we expect the system to reach a non-equilibrium steady state with a nonzero mass current flowing from the boundary with the higher mass to the boundary with the lower mass. In this paper we are interested {in understanding} the properties of this steady state distribution by computing various multi-site and multi-order correlations of the masses $\{g_i\}$ of the form 
	\begin{align}
	\mathcal{C}_{(i_1,i_2,..,i_k)}^{(m_1,m_2,...,m_k)}         = \langle g_{i_1}^{m_1}g_{i_2}^{m_2}...g_{i_k}^{m_k}\rangle_c, \label{ms-mo-corr}
	\end{align}
	 with $k \in \{1,2,3...,N$\},  $i_k \in \{1,2,...,N\}$ and $m_k \in \{1,2,3,...\}$. Here the subscript $c$ {stands for} the connected correlation {(cumulant)}. We call $\sum_i m_i = M$ as the \textit{order} of the cumulant. Our aim is to compute these correlations for large $N$.  We show that cumulants have a scaling expansions in terms of the continuum limit of the process. To demonstrate our results, we find it convenient to first discuss the scaling structure of  connected correlations up to order $M=3$, for large $N$ and then we present general case of arbitrary order with multiple sites.   	 
	 
	 For example, we show that connected correlations up to $M=3$ have the following scaling forms for large $N$ (see also 
	appendix \ref{hierarchy}):
	\begin{itemize}
		\item Correlations of order $M=1$: 
		\begin{align}
		\mathcal{C}_{i}^{(1)}=\langle g_i \rangle_c \approx  C_0^{(1)}(x),~~~\text{where}~~ x=\frac{i}{N+1}. \label{av-m-Sc}
		\end{align}
		\item Correlations of order $M=2$: 
		\begin{align}
		\begin{split}
		\mathcal{C}_{ij}^{(1,1)}=\langle g_ig_j \rangle_c &\approx \frac{1}{N}~C_1^{(1,1)}(x,y), \\
		\mathcal{C}_{ii}^{(1,1)}=\langle g_i^2 \rangle_c &\approx C_0^{(2)}(x) + \frac{1}{N}~C_1^{(2)}(x) 
		\end{split}
		,~~~\text{where}~~ x=\frac{i}{N+1},~y=\frac{j}{N+1},
		\label{av-C_ij-Sc}
		\end{align}
		\item Correlations of order $M=3$:  
		\begin{align}
		\begin{split}
		\mathcal{C}_{ijk}^{(1,1,1)}=\langle g_ig_j g_k\rangle_c &\approx \frac{1}{N^2}~C_2^{(1,1,1)}(x,y,z),  \\
		\mathcal{C}_{iik}^{(1,1,1)}=\langle g_i^2g_k \rangle_c &\approx \frac{1}{N}~C_1^{(2,1)}(x,y) +  \frac{1}{N^2}~C_2^{(2,1)}(x,y), \\
		\mathcal{C}_{iii}^{(1,1,1)}=\langle g_i^3 \rangle_c &\approx C_0^{(3)}(x)+\frac{1}{N}~C_1^{(3)}(x) +  \frac{1}{N^2}~C_2^{(3)}(x), 
		\end{split}
		\label{av-C_ijk-Sc} \\
		&~~~~~~\text{where}~~ x=\frac{i}{N+1},~y=\frac{j}{N+1},~~\text{and}~~z=\frac{k}{N+1}.
		\end{align}
	\end{itemize}
	We then show that there exists in the RATP an elegant recursive operator structure, that allows us to determine the terms in the scaling expansion above in terms of lower-order scaled cumulants. For the scaling functions defined above this recursive structure for the scaling functions {gives the results}
	\begin{align}
	C_0^{(1)}(x) &= g_L +(g_R-g_L)~x, \label{av-mass-prof}\\
	C_0^{(2)}(x) &= \frac{\mu_2}{\mu_1-\mu_2}~[C_0^{(1)}(x)]^2 \label{sc-f-C_0^2}\\
	C_1^{(2)}(x) &=\frac{\mu_1}{\mu_1-\mu_2}C_1^{(1,1)}(x,x) \label{sc-f-C_1^2}\\
	C_1^{(2,1)}(x,y) &=  \frac{2\mu_2}{\mu_1-\mu_2}~C_0^{(1)}(x)~C_1^{(1,1)}(x,y) \\
	C_2^{(2,1)}(x,y) &=\frac{\mu_1}{\mu_1-\mu_2}C_2^{(1,1,1)}(x,x,y) \\
	C_0^{(3)}(x) &= \frac{2\mu_2^2}{(\mu_1-\mu_2)^2}~[C_0^{(1)}(x)]^3 \\
	C_1^{(3)}(x) &=\frac{6\mu_1\mu_2}{(\mu_1-\mu_2)^2}~C_0^{(1)}(x)~C_1^{(1,1)}(x,x) \\
	C_2^{(3)}(x) &=\frac{\mu_1(\mu_1+\mu_2)}{(\mu_1-\mu_2)^2}~C_1^{(1,1,1)}(x,x,x),
	\end{align}
	{where }$\mu_k=\int_0^1 d\eta~\eta^k~R(\eta)$ { are the moments of the distribution $R(\eta)$.} $C_0^{(1)}(x)$ represents the average mass profile, {which is seen to be linearly interpolating between the densities of the reservoirs of the two ends.}
	
    We now state the results in a more general form. Recall that $\sum_i m_i = M$ is the order of the cumulant defined in Eq.~ \eqref{ms-mo-corr}. A cumulant of order $k$ involving $k$ distinct sites is what we call the ``max-site" correlation function. For $k=2$, the max-site cumulant is the correlation funtion $\la g_i g_k\ra_c$ with $i\neq j$. We demonstrate that for large $N$ {all cumulants have} the following expansion in scaling forms, {in orders of $1/N$},
	\begin{align}
	\mathcal{C}_{\bf i}^{\bf m} &= \langle g_{i_1}^{m_1}g_{i_2}^{m_2}...g_{i_k}^{m_k}\rangle_c \approx \frac{1}{N^{k-1}} C_{k-1}^{\bf m}({\bf x}) + \frac{1}{N^{k}} C_{k}^{\bf m}({\bf x}) + ...+ \frac{1}{N^{M-1}} C_{M-1}^{\bf m}({\bf x}),~~~\text{with}~~x_j=\frac{i_j}{N+1}, \label{scaling-form}\\
	\text{where}, & ~~{\bf i}= (i_1,i_2,..,i_k),~~{\bf m}=(m_1,m_2,...,m_k),~~~\text{and}~~{\bf x}=(x_1,x_2,..,x_k).
	\end{align}
	Here $C_l^{\bf m}({\bf x})$ is the scaling correlation function of continuous variables $x_j\in[0,1];~j=1,2,..,k$ at order $1/N^l$. {We show that all higher order ($M>k$) connected correlations of $k$ sites can be expressed in terms of the max-site cumulants with $n\leq k$ sites. We show how to generalise this to compute higher order multi-site correlations.}
	
	Our main purpose in this paper is to describe an operator recursion method by which one can compute the above scaling correlation functions systematically for different values of $M$, in terms of the 'max-site'  scaling correlation functions of the type $C^{(1,1,...,1)}_{\ell-1}(x_1,x_2,...,x_\ell)$ 
	with $\ell=1,2,...,M$.
	
	Using the same operator recursion along with the evolution equation eqn \eqref{eq:Pt} we also show that  the ``max-site" scaled correlation functions satisfy Poisson equations inside unit hypercube of dimension $\ell$ with source ``charges'' distributed appropriately,
	\begin{align}
	\left(\partial_{x_1}^2 +\partial_{x_1}^2+...+\partial_{x_\ell}^2\right)C_{\ell-1}^{(1,1,...,1)}(x_1,x_2,...,x_\ell) = S_{\ell-1}(x_1,x_2,...,x_\ell),~~\text{for}~~0 \leq x_i \leq 1,~~i=1,2,..,\ell, \label{Poisson-ell}
	\end{align}  
	with boundary conditions 
	\begin{align}
	C_{\ell-1}^{(1,1,...,1)}(x_1,x_2,...,x_j,...x_\ell)|_{x_j=0,~\text{or}~1} =0,~~\text{for}~~j=1,2,...,\ell. \label{BC-poisson-ell}
	\end{align}
	The source term $S_{\ell-1}(x_1,x_2,...,x_\ell)$ depends on the lower order max-site scaling correlation functions \emph{i.e.} 
	on  $C^{(1,1,...,1)}_{l-1}(x_1,x_2,...,x_l)$ for $0 \leq l \leq  \ell-1$.
	
	The paper is organised as follows. In section III, we show using a direct approach for the open RATP system that cumulants of order $2$ can be derived using microscopic methods. However, this approach becomes cumbersome for cumulants of higher orders, and in section IV, we describe an operator expansion method that allows us to write higher-order cumulants in terms of lower-order ones. In section V, we test the results of the operator method against simulations. In section VI, we move on to deriving equations for the max-site correlation functions, which cannot be reduced further using the operator method.  These scaled correlation functions obey Poisson equations, which we derive for the two-site and three-site correlation functions, showing that the operator method highly simplifies the derivation of these correlation functions. In section VII, we conclude with some discussion on future directions.
	
	\section{A direct approach}
	\label{direct-approach}
	 In the steady state, putting $d P(\{g_i\})/dt=0$, we find that  the master equation \eqref{eq:Pt} becomes 
	\beqa
	&& \left[ \sum_{i=1}^{N-1} \int_{0}^{\infty} dg_i' \int_{0}^{\infty} dg_{i+1}' \int_{0}^{1} d\eta R(\eta) P(\{g_i'\}) \right. \nonumber \\ 
	&&\times  \{\delta(g_i - g_i' + \eta g_i')\delta(g_{i+1}- g_{i+1}' - \eta g_i) + \delta(g_{i+1} - g_{i+1}' + \eta g_{i+1}')\delta(g_{i}- g_{i}' - \eta g_{i+1})\} \nonumber\\ 
	&&+ \int_{0}^{\infty} dg_1' \int_{0}^{\infty} dG_L \int_{0}^{1} d\eta R(\eta) P(\{g_i'\}) P_L(G_L) \{\delta(g_1 - g_1' + \eta g_1') + \delta(g_1 - g_1' - \eta G_L)\} \nonumber\\ 
	&&+ \left. \int_{0}^{\infty} dg_N' \int_{0}^{\infty} dG_R \int_{0}^{1} d\eta R(\eta) P(\{g_i'\}) P_R(G_R) \{\delta(g_N - g_N' + \eta g_N') + \delta(g_N - g_N' - \eta G_R)\} \right]=0
	\label{eq:Pss}
	\eeqa
	From this equation one can get the equations satisfied by the correlations functions at different order $M$. For example, at order $M=1$ one obtains 
	\begin{align}
	\langle g_{i+1} \rangle_c -2 \langle g_{i} \rangle_c +\langle g_{i-1} \rangle_c=0,~~\text{for}~~i=1,2,..,N,~~\text{with},~~
	\langle g_0\rangle=\langle G_L \rangle =g_L,~~\text{and}~~\langle g_{N+1}\rangle=\langle G_R \rangle =g_R.
	\end{align} 
	This is a simple equation and one can solve this equation exactly. The solution is given by 
	\begin{align}
	\mathcal{C}_{i}^{(1)}=\langle g_{i} \rangle_c = g_L + (g_R-g_L)~\frac{i}{N+1},~~\text{for}~~i=1,2,..,N. \label{av-m-ex-sol}
	\end{align}	
	In a similar way one can get the equations satisfied by the two-point correlations that appear at order $M=2$. Starting again from Eq.~\eqref{eq:Pss}, one computes $\mathcal{C}_{i,j}^{(1,1)}=\langle g_{i} g_j\rangle_c$ and obtains the following equations 
	\begin{align}
	\mu_1&\left[ \mathcal{C}_{i+1,j}^{(1,1)}+\mathcal{C}_{i-1,j}^{(1,1)}+\mathcal{C}_{i,j+1}^{(1,1)}+\mathcal{C}_{i,j-1}^{(1,1)} -4 \mathcal{C}_{i,j}^{(1,1)} \right] =0,~~\text{for},~~1\leq (i,j) \leq N,~~\text{and}~~|i-j|\geq 2, \\
	\mu_1&\left[ \mathcal{C}_{i+1,i+1}^{(1,1)}+\mathcal{C}_{i-1,i+1}^{(1,1)}+\mathcal{C}_{i,i+2}^{(1,1)}+\mathcal{C}_{i,i}^{(1,1)} -4 \mathcal{C}_{i,i+1}^{(1,1)}\right]-\mu_2\left[ \mathcal{C}_{i,i}^{(1,1)}+\mathcal{C}_{i+1,i+1}^{(1,1)}\right] = \mu_2\left[ (\mathcal{C}_{i}^{(1)})^2 +(\mathcal{C}_{i+1}^{(1)})^2\right], \\ 
	&~~~~~~~~~~~~~~~~~~~~~~~~~~~~~~~~~~~~~~~~~~~~~~~~~~~~~~~~~~~~~~~~~~~~~~~~~~~~~~~
	~~~~~~~~~~~~~~~~~~~~~~~~~~~
	~~\text{for},~~1\leq i \leq N-1, \nonumber \\
	\mu_1&\left[ \mathcal{C}_{i,i+1}^{(1,1)}+ \mathcal{C}_{i+1,i}^{(1,1)}+\mathcal{C}_{i-1,i}^{(1,1)} +\mathcal{C}_{i,i-1}^{(1,1)}-4 \mathcal{C}_{i,i}^{(1,1)}\right]+\mu_2\left[ \mathcal{C}_{i-1,i-1}^{(1,1)}+\mathcal{C}_{i+1,i+1}^{(1,1)}+2\mathcal{C}_{i,i}^{(1,1)}\right] \nonumber \\ 
	&~~~~~~~~~~~~~~~~~~~~~~~~~~~~~~~~~~~~~
	= -\mu_2\left[ (\mathcal{C}_{i-1}^{(1)})^2 +(\mathcal{C}_{i+1}^{(1)})^2 +2(\mathcal{C}_{i}^{(1)})^2 \right]  
	~~~\text{for},~~1\leq i \leq N, \nonumber
	\end{align}
	with boundary conditions $\mathcal{C}_{0,j}^{(1,1)}=\mathcal{C}_{N+1,j}^{(1,1)}=\mathcal{C}_{j,0}^{(1,1)}=\mathcal{C}_{N+1,j}^{(1,1)}=0$ for $1\leq j \leq N$. Note that the equations for two-point functions closes into themselves as they do not involve higher order or higher point correlations. Also remember that $\mathcal{C}_{i}^{(1)}= \langle  g_i \rangle$ is the average mass at site $i$. One can solve the above equations exactly in this case also and the solutions are given by 
	\begin{align}
	\mathcal{C}_{i,j}^{(1,1)}&= \begin{cases}
	\begin{split}
	&A~\frac{(g_R-g_L)^2}{N+1}~\frac{i}{N+1}\left(1-\frac{j}{N+1} \right),~~\text{for}~~1\leq i < j\leq N, \\
	&A~\frac{(g_R-g_L)^2}{N+1}~\frac{j}{N+1}\left(1-\frac{i}{N+1} \right),~~\text{for}~~1\leq j < i\leq N, \\
	\end{split},~~~~~\text{where}~~A= \frac{\mu_2}{\mu_1-\mu_2},\\
	&A~\frac{(g_R-g_L)^2}{N+1}~\frac{\mu_1}{\mu_1-\mu_2}~\frac{i}{N+1}\left(1-\frac{i}{N+1} \right) + \frac{\mu_2}{\mu_1-\mu_2}~(\mathcal{C}_{i}^{(1)})^2-\frac{\mu_1\mu_2^3}{2(\mu_1-\mu_2)^4}\frac{(g_L-g_R)^2}{(N+1)^3}~~\text{for}~~1\leq i \leq N,~i=j,
	\end{cases}
	\label{C_ij-sol}
	\end{align}
	where $\mathcal{C}_{i}^{(1)}$ is provided in Eq.~\eqref{av-m-ex-sol}.
	Notice that the solutions in Eqs.~\eqref{av-m-ex-sol} and \eqref{C_ij-sol} are in the scaling forms as in Eqs.~\eqref{av-m-Sc} and \eqref{av-C_ij-Sc} with the scaling functions given explicitly as  
	\begin{align}
	C_0^{(1)}(x) &= g_L +(g_R-g_L)~x  \label{sol-C_0^(1)}\\
        \begin{split}
	C_1^{(1,1)}(x,y)&=\begin{cases}
	&\frac{\mu_2}{\mu_1-\mu_2}(g_R-g_L)^2~x(1-y),~~~~~~~~\text{for}~~0<x<y<1 \\
	&\frac{\mu_2}{\mu_1-\mu_2}(g_R-g_L)^2~y(1-x),~~~~~~~~\text{for}~~0<y<x<1
	\end{cases}  \\
	C_0^{(2)}(x) &= \frac{\mu_2}{\mu_1-\mu_2}~[g_L +(g_R-g_L)~x]^2,~~\text{for}~~0<x<1~~\text{and}, \\
	C_1^{(2)}(x) &=\frac{\mu_1}{\mu_1-\mu_2}~(g_R-g_L)^2~x(1-x),~~\text{for}~~0<x<1.
	\end{split}
	\label{explicit-scaling-functions-C_ij}
	\end{align}
Let us now look at correlations of order $M=3$. At this order the correlation that involves maximum three distinct sites is 
$\mathcal{C}_{i,j,k}^{(1,1,1)}=\langle g_ig_jg_k\rangle_c$ ({with} $i\neq j\neq k$). Once again starting from the steady state master equation \eqref{eq:Pss}, one can write the equations satisfied by this correlation as well as other two  correlations $\mathcal{C}_{i,i,k}^{(1,1,1)}$ and $\mathcal{C}_{i,i,i}^{(1,1,1)}$ at order $M=3$, which once again close into themselves as their equations do not involve higher order correlations. This closing property can be observed at every order $M$ of correlations. This is due to linear and locally independent properties of the mass mixing process that defines the RATP. One can try to solve these equations at every order microscopically as done for $M=1$ and $M=2$. But it can be easily realised that the microscopic procedure soon becomes quite cumbersome as the order of the correlations $M$ increases. 

{Instead, in the next section, we present a new method to compute} solutions in the scaling form assuming that scaling similar to that shown for orders $M=1$ and $2$ holds at every order. The assumption of the existence of scaling structure as in Eq.~\eqref{scaling-form} at orders $M>2$ can be observed numerically{, and} we later present numerical support {for this} assumption. To find the correlations in the scaling forms we, in the next section, employ a novel operator structure.	
	\section{An operator expansion method} 
	\label{sec:operator}

\noindent	
In the previous section we have shown that the correlations at order $M=1$ and $2$ have expansions in powers of $1/N$ where coefficients have well defined scaling forms in Eqs.~\eqref{av-m-Sc}, \eqref{av-C_ij-Sc} with scaling functions given explicitly in Eq.~\eqref{explicit-scaling-functions-C_ij}. We have obtained these scaling functions from the exact microscopic solutions. In this section we will discuss an alternate method through which one can compute all the cumulants upto a given order $M=n$, given that all possible max-site scaling correlation functions are known upto order $n$. For example, if  $C_0^{(1)}(x),~C_1^{(1,1)}(x,y)$ and $C_2^{(1,1,1)}(x,y,z)$ are known, then all other cumulants upto order $M=3$ can be computed. This is because higher order cumulants in the steady state can be expressed in terms of lower order cumulants by breaking the cumulants in disconnected cumulants of lower orders. This procedure can be elegantly expressed in terms of operator recursion relations which we derive below. 
	
	Let us first focus on the second order ($M=2$) cumulants at a single site $\langle g_i^2 \ra$. We consider the evolution of the correlation function $\la g_i g_{i+1}\ra$ which can be easily obtained from the time dependent master equation \eqref{eq:Pt} as
	\beqa
	\frac{d}{dt} \la g_i g_{i+1} \ra = &\frac{1}{2}&\mu_1 \left( \la g_i^2 \ra + \la g_{i+1}^2 \ra + \la g_{i-1} g_{i+1} \ra + \la g_i g_{i+2} \ra - 4 \la g_i g_{i+1} \ra \right) - \frac{1}{2} \mu_2( \la g_i^2 \ra + \la g_{i+1}^2 \ra )
	\eeqa
	Since the LHS is zero in the steady-state we have 
	\begin{align}
	\la g_i^2 \ra + \la g_{i+1}^2 \ra = \frac{\mu_1}{\mu_1-\mu_2} (4 \la g_ig_{i+1}\ra - \la g_{i-1} g_{i+1} \ra - \la g_i g_{i+2} \ra ).
	\end{align}
	Now using the scaling forms in Eqs.~\eqref{av-m-Sc} and \eqref{av-C_ij-Sc} in the above equations it is easy to see that, in the large $N$ limit,  $\la g_i^2 \ra_c \approx$$\frac{\mu_2}{\mu_1-\mu_2}(C_0^{(1)})^2$ at order $N^0$, while at order $N^{-1}$, it is equal to $\frac{\mu_1}{\mu_1-\mu_2}C_1^{(1,1)}(x,y)|_{y \to x}$. {This suggests the} following operator recursion relation
	\begin{align}
	g_i^2 \equiv \frac{\mu_1}{\mu_1-\mu_2}~g_i . g_i,~~\text{where}~~g_i.g_i = \lim \limits_{j \to i} g_ig_j. \label{op-eq-o-2}
	\end{align}
	As will soon become clear, this operator method turns out to be a very efficient method to compute cumulants at different orders.
	The meaning of this operator equation is best understood by specifying its action under steady state averages. Inside the angular brackets {the ``separated" product operator,} $\la g_i.g_i\ra_c$ {gives the} scaling correlation function $C_1^{(1,1)}(x,y)|_{y \to x}$ at order $1/N$. Let us now demonstrate how this operator equation {is applied to obtain} $\la g_i^2 \ra_c$. bluew{Taking a} steady state average on both sides of Eq.~\eqref{op-eq-o-2}.
	\begin{align}
	\la g_i^2 \ra = \frac{\mu_1}{\mu_1-\mu_2}~\la g_i . g_i \ra =  \frac{\mu_1}{\mu_1-\mu_2}~\lim \limits_{j \to i}\la g_i g_j \ra 
	\end{align}
	Here the limit $j \to i$ is understood in the sense that the corresponding scaling variables approach each other \emph{i.e.} $y \to x$ in the large $N$ limit where $x=\frac{i}{N+1}$ and $y=\frac{j}{N+1}$.
	Now expand the averages in terms of the cumulants and connected correlation functions as 
	\begin{align}
	\la g_i \ra^2 + \la g_i^2 \ra_c &= \frac{\mu_1}{\mu_1-\mu_2}~\lim \limits_{j \to i} \left[ \la g_i\ra \la g_j \ra + \la g_i g_j \ra_c\right] \\
	\la g_i^2 \ra_c &=  \frac{\mu_2}{\mu_1-\mu_2} \la g_i \ra^2 + \frac{\mu_1}{\mu_1-\mu_2}~\lim \limits_{j \to i} \left[  \la g_i g_j \ra_c\right] \\
	\la g_i^2 \ra_c &=  \frac{\mu_2}{\mu_1-\mu_2} \la g_i \ra^2 + \frac{1}{N}~\frac{\mu_1}{\mu_1-\mu_2}~\lim \limits_{y \to x}  \left[C_1^{(1,1)}(x,y)\right] \\
	\la g_i^2 \ra_c &=  \frac{\mu_2}{\mu_1-\mu_2} (C_0^{(1)}(x))^2 + \frac{1}{N}~\frac{\mu_1}{\mu_1-\mu_2}~C_1^{(1,1)}(x,x)
	\label{var-g_i}
	\end{align}
	as announced in Eq.~\eqref{av-C_ij-Sc}. 
	Because of the locality of the evolution equations for the RATP, the recursion equation \eqref{op-eq-o-2} can be used to compute higher order and multiple site cumulants of the form $\la g_i^2... \ra$ involving $g_i^2$.
	{We now} demonstrate {the use of} this procedure to derive bluew{various} other cumulants: 
	\begin{itemize}
		\item Computation of $\la g_i^2 g_k \ra_c$ where $k \neq i$:
		\begin{align}
		&\la g_i^2g_k \ra = \frac{\mu_1}{\mu_1-\mu_2}~\la g_i.g_i g_k \ra= \frac{\mu_1}{\mu_1-\mu_2}~
		\lim \limits_{j \to i}~\la g_i g_j g_k \ra, 
		\end{align}
		Expanding both sides in cumulants, we find
		\begin{align}
		&\la g_i^2g_k \ra = \frac{\mu_1}{\mu_1-\mu_2}~\la g_i.g_i g_k \ra, \\
		&\la g_i^2g_k \ra_c + 2 \la g_i \ra \la g_i g_k \ra_c + \la g_i^2 \ra_c \la g_k \ra + \la g_i \ra^2 \la g_k \ra
		= \frac{\mu_1}{\mu_1-\mu_2}~\lim \limits_{j \to i}\left[ \la g_i\ra^2\la g_k \ra + \la g_k \ra \la g_i g_j \ra_c +  \la g_i \ra \la g_j g_k \ra_c \right. \nonumber \\ 
		&~~~~~~~~~~~~~~~~~~~~~~~~~~~~~~~~~~~~~~~~~~~~~~~~
		~~~~~~~~~~~~~~~~~~~~~~~~~~~~~~~~~~~~~~~~~~~~~~~~~~
		\left. +\la g_j \ra \la g_i g_k \ra_c + \la g_i g_j g_k \ra_c\right], \\
		&\la g_i^2g_k \ra_c = \frac{1}{N}~\frac{2 \mu_2}{\mu_1 - \mu_2}~C_0^{(1)}(x)C_1^{(1,1)}(x,z) + \frac{1}{N^2}~\frac{\mu_1}{\mu_1 -\mu_2}~C_2^{(1,1,1)}(x,x,z),~~\text{as~stated~in~Eq.~\eqref{av-C_ijk-Sc}}.
		\label{Cc-g_i-sq-g_k}
		\end{align}

		\item Computation of $\la g_i^2 g_k^2 \ra_c$ where $k \neq i$: 
		\begin{align}
		&\la g_i^2g_k^2\ra = \frac{\mu_1^2}{(\mu_1-\mu_2)^2} \la g_i.g_ig_k.g_k \ra,~~~\text{expanding~both~sides~in~connected~correlations~we~get,} \\
		&\la g_i^2 g_k^2\ra_c + 2 \la g_i\ra \la g_i g_k^2 \ra_c +2 \la g_k \ra \la g_i^2 g_k\ra_c + \la g_i^2 \ra_c \la g_k^2 \ra_c + 2 \la g_ig_k\ra_c^2 
		+ \la g_i^2\ra_c \la g_k\ra^2 + \la g_i\ra^2 \la g_k\ra_c  + 4 \la g_i\ra \la g_k \ra \la g_i g_k \ra_c\nonumber \\ 
		& ~~~+ \la g_i \ra^2 \la g_k\ra ^2 
		=  \frac{\mu_1^2}{(\mu_1-\mu_2)^2}~\lim \limits_{\substack{j\to i \\ l \to k}} \left [ \la g_ig_j g_k g_l \ra_c + \la g_i\ra \la g_jg_kg_l \ra_c + \la g_j \ra \la g_i g_k g_l\ra_c +\la g_k \ra \la g_i g_j g_l\ra_c+\la g_l \ra \la g_i g_j g_k\ra_c \right. 
		\nonumber \\
		&+\la g_i g_j\ra_c \la g_k g_l\ra_c +\la g_i g_k\ra_c \la g_j g_l\ra_c +\la g_i g_l\ra_c \la g_j g_k\ra_c + \la g_i g_j\ra_c \la g_k \ra \la g_l\ra + \la g_i g_k\ra_c \la g_j \ra \la g_l\ra + \la g_i g_l\ra_c \la g_j \ra \la g_k\ra 
		\nonumber \\
		& + \la g_j g_k\ra_c \la g_i \ra \la g_l\ra + \la g_j g_l\ra_c \la g_i \ra \la g_k\ra + \la g_k g_l\ra_c \la g_i \ra \la g_j\ra + \la g_j \ra \la g_l\ra \la g_i \ra \la g_k\ra. \label{av-g_i-sq-g_k-sq}
		\end{align}
		Now we use the scaling forms of $\la g_i \ra$, $\la g_i g_j\ra_c$, $\la g_i^2 \ra_c$, $\la g_i^2 g_k\ra_c$ given in Eqs.~\eqref{av-m-Sc}, \eqref{av-C_ij-Sc} and \eqref{Cc-g_i-sq-g_k} and, the following max-site correlations 
		\begin{align}
		\la g_i g_j g_k \ra_c &= \frac{1}{N^2}~C_2^{(1,1,1)}(x,y,z),~~\text{for}~i \neq j\neq k,~~\text{with}~x=\frac{i}{N+1},~y=\frac{j}{N+1},~z=\frac{k}{N+1},~\text{and} \nonumber \\
		\la g_i g_j g_k g_l \ra_c &=\frac{1}{N^3}~C_3^{(1,1,1,1)}(x,y,z,w),~~\text{for}~i \neq j\neq k \neq l,~~\text{additionally~with}~w=\frac{l}{N+1},
		\end{align}
		on both sides of Eq.~\eqref{av-g_i-sq-g_k-sq}. After carrying out some straightforward manipulations and simplifications, we get 
		
			\begin{align}
			\begin{split}
			\la g_i^2 g_k^2 \ra_c &\approx \frac{1}{N}~\frac{4 \mu_2^2}{(\mu_1-\mu_2)^2}~C_0^{(1)}(x)~C_0^{(1)}(z)~C_1^{(1,1)}(x,z)  \\ 
			&~~~~~~~+ \frac{1}{N^2} \frac{2 \mu_1 \mu_2}{(\mu_1-\mu_2)^2}~\left[ C_0^{(1)}(x)C_2^{(1,1,1)}(x,z,z)+C_0^{(1)}(z)C_2^{(1,1,1)}(x,x,z)\right] \\
			&~~~~~~~~~~~~~~~~~~~~~~~+ \frac{1}{N^3}~\frac{\mu_1^2}{(\mu_1-\mu_2)^2}~C_3^{(1,1,1,1)}(x,x,z,z)
			\end{split}
			\label{C_1^(22)}
			\end{align}

		\item Generalisation to higher order and higher point correlations of the form $\la g_i^2 g_{k_1}...g_{k_n} \ra$ where none of the $k_j \neq i$ for all $j=1,2,..,n$.
		\begin{align}
		&\la g_i^2g_{k_1}...g_{k_n}  \ra = \frac{\mu_1}{\mu_1-\mu_2}~\la g_i.g_i g_{k_1}...g_{k_n}  \ra, \\
		&\la g_i^2g_{k_1}...g_{k_n}  \ra_c + 2 \la g_i \ra \la g_i g_{k_1}...g_{k_n}  \ra_c + \la g_i^2 \ra_c \la g_{k_1}...g_{k_n}  \ra_c + \la g_i \ra^2 \la g_{k_1}...g_{k_n}  \ra_c  \\
		&= \frac{\mu_1}{\mu_1-\mu_2}~\lim \limits_{j \to i}\left[ \la g_i\ra^2\la g_{k_1}...g_{k_n}  \ra_c + \la g_{k_1}...g_{k_n}  \ra_c \la g_i g_j \ra_c +  \la g_i \ra \la g_j g_{k_1}...g_{k_n}  \ra_c +\la g_j \ra \la g_i g_{k_1}...g_{k_n}  \ra_c + \la g_i g_j g_{k_1}...g_{k_n}  \ra_c\right]. \nonumber 
		\end{align}
	\end{itemize}
	Note that the operator equation \eqref{op-eq-o-2} can not be used to compute correlations which involve $g_i^3$, for example, 
	$\la g_i^3 \ra_c,~\la g_i^3g_k \ra_c$ etc. For that we need operator recursion equation for $g_i^3$. Following the same procedure as was used to derive Eq.~\eqref{op-eq-o-2}, we find
		\begin{align}
		g_i^3 &= \frac{\mu_1+\mu_2}{\mu_1-\mu_2} g_i^2 \cdot g_i.  \label{op-eq-o-3}
		\end{align}
	Using this operator equation one can straightforwardly workout the following cumulants involving $g_i^3$
	\begin{itemize}
		\item $\la g_i^3\ra_c$ : 
		\begin{align}
		&\la g_i^3 \ra = \frac{\mu_1+\mu_2}{\mu_1-\mu_2}~\la g_i^2.g_i \ra, ~~~\text{expanding~both~sides~in~connected~correlations~we~get,} \\
		&\la g_i^3\ra_c + 3 \la g_i\ra \la g_i^2 \ra_c + \la g_i \ra^3 = \frac{\mu_1+\mu_2}{\mu_1-\mu_2}~\lim \limits_{j \to i}\left[ \la g_i\ra^2 \la g_j \ra + 2 \la g_i \ra \la g_i g_j \ra_c + \la g_i^2 \ra_c \la g_j \ra + \la g_i^2 g_j \ra_c \right] 
		\end{align}
		Now once again we  use the scaling forms $\la g_i \ra$, $\la g_i g_j\ra_c$, $\la g_i^2 \ra_c$, $\la g_i^2 g_k\ra_c$ given in Eqs.~\eqref{av-m-Sc}, \eqref{av-C_ij-Sc} and \eqref{Cc-g_i-sq-g_k} on both sides of the above equation. After simplifying we get 
		\begin{align}
		\la g_i^3 \ra_c = \frac{2 \mu_2^2}{(\mu_1-\mu_2)^2}~(C_o^{(1)}(x))^3 + \frac{1}{N}~\frac{\mu_1\mu_2}{(\mu_1-\mu_2)^2} C_0^{(1)}(x)~C_1^{(1,1)}(x,x)+\frac{1}{N^2}~\frac{\mu_1(\mu_1+\mu_2)}{(\mu_1-\mu_2)^2}~C_2^{(1,1,1)}(x,x,x)
		\label{g_i^3_c}
		\end{align}
		
		\item $\la g_i^3 g_k \ra_c $ for $k \neq i$: 
		\beqa
		\la g_i^3 g_k \ra = \frac{\mu_1(\mu_1+\mu_2)}{(\mu_1-\mu_2)^2} \la g_i \cdot g_i \cdot g_i \cdot g_k \ra
		\eeqa
		Expanding both sides in cumulants, and using the scaling forms for $\la g_i \ra$, $\la g_i^2 \ra_c$, $\la g_i^2 g_k \ra_c$, $\la g_i^3 \ra_c$ as derived above, we get
		\beqa
		\la g_i^3 g_k \ra &=& \frac{1}{N} \frac{6 \mu_2^2 - \mu_1 \mu_2}{(\mu_1-\mu_2)^2} C_0^{(1)}(x) C_0^{(1)}(y) C_1^{(1,1)}(x,x) + \frac{1}{N} \frac{12 \mu_1 \mu_2 - 6 \mu_2^2}{(\mu_1-\mu_2)^2} C_0^{(1)}(x)^2 C_1^{(1,1)}(x,y) \nonumber\\
		& &+ \frac{1}{N^2} \frac{6 \mu_2^2}{(\mu_1-\mu_2)^2} C_1^{(1,1)}(x,x) C_1^{(1,1)}(x,y) 
		+ \frac{1}{N^2} \frac{6 \mu_2^2}{(\mu_1-\mu_2)^2} C_0^{(1)}(x) C_2^{(1,1,1)}(x,x,y) \nonumber\\
		& &+ \frac{1}{N^3} \frac{\mu_1(\mu_1+\mu_2)}{(\mu_1-\mu_2)^2} C_3^{(1,1,1,1)}(x,x,x,y)
		\eeqa 
	\end{itemize}
	One can generalize the above operator method to compute cumulants of arbitrary powers using the following  operator recursion relation {for $g_n$}
	 $g_i^n$ 
		\begin{align}
		g^{n}_i = \frac{1}{\la 1-(1-\eta)^k\ra - \mu_n}\sum_{k=1}^{n-1} \binom{n}{k} \mu_k~g^{n-k}_i \cdot g^k_i,~~
		n=1,2,.. \label{op-eq-o-n}
		\end{align}
	which is  derived in the appendix~\ref{op-gen-n}.
	
	\subsection{General equations for the higher order cumulants}
\label{gen_high_cumu}
	In this section we give  general expressions for all multi-site cumulants of higher order in terms of the max-site cumulants.
	To do this we first observe ( from previous examples) that in the operator form one can write 
	\be
	g_i^n = f_n \overbrace{g_i \cdot g_i \cdot \dots}^{n \mbox{ times}} 
	\label{op-split}
	\ee	
	where $f_n$ are constants independent of the site index, and are functions of the moments of $R(\eta)$. Inserting this form into Eq.~ \eqref{op-eq-o-n}, we get a recursion relation for $f_n$,
	\be
	f_n = \frac{1}{\la 1-(1-\eta)^k\ra - \mu_n}\sum_{k=1}^{n-1} \binom{n}{k} \mu_k~ f_{n-k} f_k \label{fn-recur}
	\ee
	This equation can be solved in closed form for certain special cases of $R(\eta)$ e.g. uniform distribution.  However for arbitrary $R(\eta)$, this can be solved recursively to determine $f_n$ for all orders. It also follows that the $f_n$s up to order $n$ are functions only of the first $n$ moments of $R(\eta)$. {The first few $f_n$s} are explicitly given by
\bea
f_1&=&1, \nonumber \\
f_2&=& \frac{\mu_1}{\mu_1-\mu_2}, \nonumber \\
f_3&=&\frac{\mu_1+\mu_2}{\mu_1-\mu_2}f_1f_2=\frac{\mu_1(\mu_1+\mu_2)}{(\mu_1-\mu_2)^2} \label{f-sol}\\
f_4&=&\frac{ \left[ 2(\mu_1+\mu_3)f_3+3\mu_2 f_2^2 \right]}{2(\mu_1+\mu_3)-3\mu_2 -\mu_4} \nonumber \\
&=&\frac{\mu_1}{(\mu_1 - \mu_2)^2} \frac{2(\mu_1+\mu_3)(\mu_1+\mu_2)+3 \mu_1\mu_2}{2(\mu_1+\mu_3)-3\mu_2 -\mu_4}\nonumber 
\eea	
Once the sequence $f_n$ is determined, all the cumulants of $g_i$ till order $n$ can be computed in terms of the lower order cumulants and $f_m$'s with $1\leq m \leq n$ and the max-site cumulants of order till $n$. To see this one needs to take expectation on both sides of Eq.~ \eqref{op-split}. On the left hand side one expands $\la g_i^n\ra$ in terms of its lower order cumulants as 
	\be
	\la g_i^n\ra = n! \sum_{\{p_m\}} \prod_{m=0}^n \frac{1}{p_m! (m!)^{p_m}} \la g_i^m\ra_c^{p_m}~\delta_{n ,\sum_{m=0}^n mp_m}
	\label{expan_g_i^n}
	\ee
	for example, $\la g_i^3\ra= \la g_i^3\ra_c + 3 \la g_i \ra \la g_i^2 \ra_c + \la g_i\ra^3 $. In the above equation $\delta_{a,b}$ represents Kroneker delta and $p_m$'s take non-negative integer values. Similarly, one can expand the right hand side 
	as 
	\bea
	\la \overbrace{g_i \cdot g_i \cdot \dots}^{n \mbox{ times}} \ra 
	&=& \lim \limits_{\substack{
             j \to i\\
           k \to j \\
           ...}}~\la g_ig_jg_k... \ra   
           = \lim \limits_{\substack{
             j \to i\\
           k \to j \\
           ...}} 
           \left[ \la g_ig_jg_k... \ra_c + \la g_i\ra \la g_j g_k... \ra + \la g_j\ra \la g_i g_k...\ra + ... + \la g_i g_j \ra_c \la g_k... \ra_c + ... \right]
           \label{expan_g_iprod}
	\eea
	Again for example,
	$\la {g_i \cdot g_i \cdot g_i} \ra = \la g_i\ra^3 + 3 \la g_i \ra \la g_i g_j\ra_c\big{|}_{j \to i} + \la g_i g_j g_k \ra_c\big{|}_{k \to j\to i}.$

Note that, Inserting the expansion in Eq.~\eqref{expan_g_i^n} on the left hand side and the expansion in Eq.~\eqref{expan_g_iprod} on the right hand side of Eq.~\eqref{op-split}, one can express $\la g_i^n\ra_c$ in terms of the  max-site correlations till order $n$. 
Now, using the fact that in the scaling limit, $\la g_i \ra \rightarrow  C_0^{(1)}(x)$, $\la g_i \cdot g_i \ra_c \rightarrow C^{(1,1)}(x,x)$,  $\la g_i \cdot g_i \cdot g_i \ra_c \rightarrow C^{(1,1,1)}(x,x,x)$, and so on, one can find the scaling form of $\la g_i^n\ra_c$. For our example of $n=3$, we find,
\bea
\la g_i^3 \ra_c &=& f_3 \left[\la g_i\ra^3 + 3 \la g_i \ra \la g_i g_j\ra_c\big{|}_{j \to i} + \la g_i g_j g_k \ra_c\big{|}_{k \to j\to i}\right] -\left[ 3 \la g_i \ra \la g_i^2 \ra_c + \la g_i\ra^3  \right]
\eea{
We first make use of  the solutions of $f_n$ from Eq.~\eqref{fn-recur} and the scaling forms from Eq.~\eqref{scaling-form}. Finally,  inserting their explicit relations derived in the previous section and performing algebraic simplifications, we get 
\bea
\la g_i^3\ra_c=\frac{2 \mu_2^2}{(\mu_1-\mu_2)^2}C_o^{(1)}(x)^3 + \frac{1}{N}~\frac{6\mu_1\mu_2}{(\mu_1-\mu_2)^2}~C_0^{(1)}(x)C_1^{(1,1)}(x,x) + \frac{1}{N^2}~\frac{\mu_1(\mu_1+\mu_2)}{(\mu_1-\mu_2)^2}~C_2^{(1,1,1)}(x,x,x)
\label{<g_i^3>}
\eea
as obtained earlier in Eq.~\eqref{g_i^3_c}. Similarly, for $n=4$ one finds
\bea
\la g_i^4\ra =f_4 \left(C_0^{(1)}(x)^4 \right. &+& \left.\frac{6}{N} C_0^{(1)}(x)^2 C_1^{(1,1)}(x,x)
+ \frac{3}{N^2} C_1^{(1,1)}(x,x)^2  \right. \nonumber \\
&+& \left.  \frac{4}{N^2} C_0^{(1)}(x)C_2^{(1,1,1)}(x,x,x) 
+ \frac{1}{N^3}C_3^{(1,1,1,1)}(x,x,x,x) \right).
\label{<g_i^4>}
\eea
Expanding the moment $\la g_i^4\ra$ on lhs in terms of the cumulants and after some algebraic manipulations one finds
\bea
\la g_i^4 \ra_c &=& \left( f_4 - \frac{6 \mu_2^2+4\mu_1\mu_2+\mu_1^2}{(\mu_1-\mu_2)^2} \right) C_0^{(1)}(x)^4 
+ \frac{1}{N}~\left(6 f_4 -\frac{24\mu_1\mu_2+6\mu_1^2}{(\mu_1-\mu_2)^2}\right) C_0^{(1)}(x)^2 C_1^{(1,1)}(x,x)
\nonumber \\
&&+ \frac{1}{N^2}~\left(3 f_4 - \frac{3 \mu_1^2}{(\mu_1-\mu_2)^2}\right) C_1^{(1,1)}(x,x)^2 
+ \frac{1}{N^2}~4\left(f_4 -\frac{\mu_1(\mu_1+\mu_2)}{(\mu_1-\mu_2)^2}\right)C_0^{(1)}(x)C_2^{(1,1,1)}(x,x,x) 
\label{g_i^4-c}\\
&&~~~~~~~~~~~~~~~~~~~~~~~~~~~~~~~~~~~~~~
+ \frac{1}{N^3}~f_4 ~C_2^{(1,1,1,1)}(x,x,x,x),\nonumber
\eea
where explicit expression of $f_4$ is given in Eq.~\eqref{f-sol}.
 Thus once again we observe that any cumulant at order $n$, can be expressed in terms of all the max-site scaling correlation functions $C_{j-1}^{(1,1,...,1)}(x_1,x_2,...,x_j)$ of order $j$ till $n$ \emph{i.e.} $\forall~j$ such that $1\leq j \leq n$. {In section \ref{sec:Poisson}, we turn to} the question is how do we find these max-site scaling correlation functions at a given order{, where we find} that at any order $n$ the max-site scaling correlation function satisfies a Poisson equation inside a unit cube of dimension $n$ with source ``charges'' distributed appropriately.
 {But in the next section, we first present numerical verification of the results of this section using Monte-Carlo simulations of the RATP.}
 
 \begin{figure}[h]
\centering
			\includegraphics[width=0.46\linewidth]{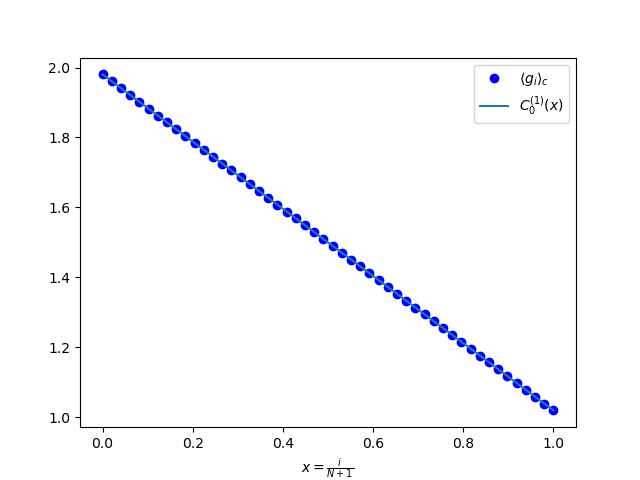}
		\caption{Comparison of simulation results (filled circles) with theory (solid lines) of the expectation value of $g_i$ on a lattice of size $N=49$.  Solid points are obtained from simulation using the jump distribution given in Eq.~\eqref{R(eta)}. The parameters used in the simulation are $g_L=2$ and $g_R=1$. }
		\label{mean}
\end{figure}

\begin{figure}[h]
\centering
		\subfigure[]{
			\includegraphics[width=0.46\linewidth]{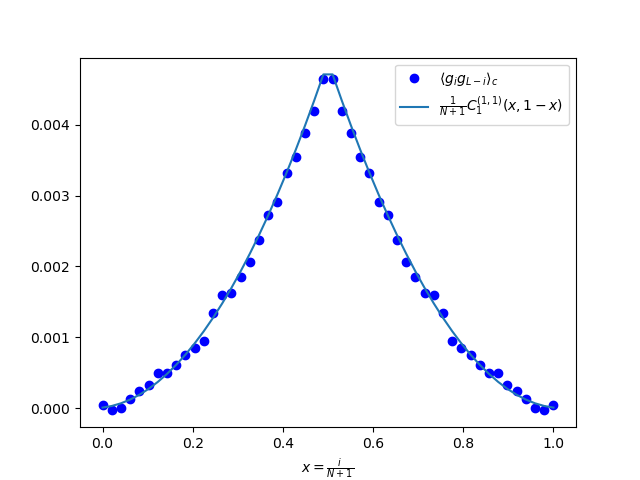}
		}
		\subfigure[]{
			\includegraphics[width=0.46\linewidth]{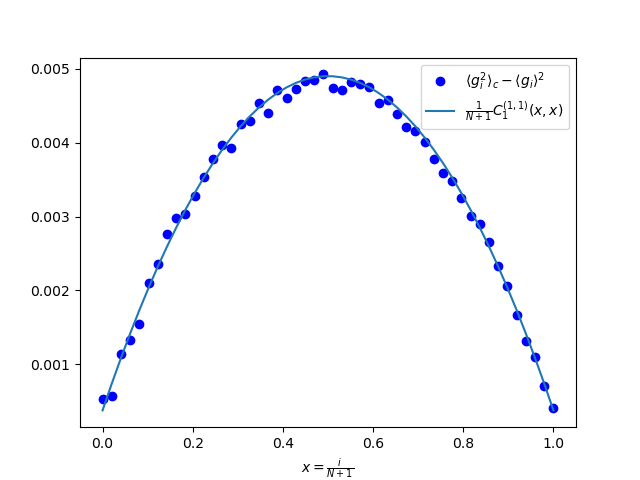}
		}
		\caption{Comparison of simulation results (filled circles) with theory (solid line) of (a) the correlation function $\la g_i g_j \ra$ along the line $j = N - i$, and (b) the cumulant $\la g_i^2 \ra_c$ {, with the $O(1)$ contribution, $\la g_i \ra^2$, subtracted}. The parameters used in the simulation are same as fig.~\ref{mean}.}
		\label{C_0^1-and-C_1^11}
\end{figure}
 
\begin{figure}[h]
\centering	
		\subfigure[]{
	\includegraphics[width=0.46\linewidth]{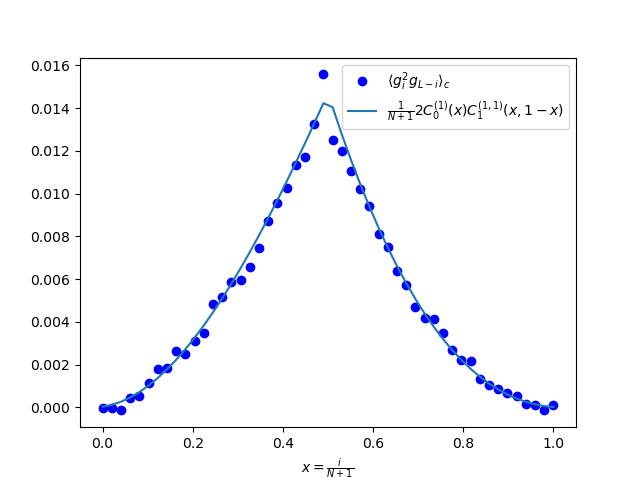}
}
\subfigure[]{
	\includegraphics[width=0.46\linewidth]{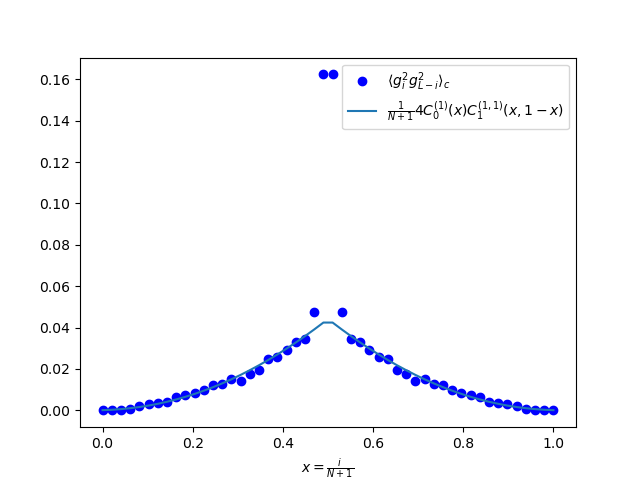}
}
\caption{Comparison of simulation results (filled circles) with theory (solid line) of the two-site cumulants (a) $\la g_i^2 g_j \ra_c$ and (b) $\la g_i^2 g_j^2 \ra_c$, both along the line $j = N - i$. The parameters used in the simulation are same as fig.~\ref{mean}. The deviation from the scaling behavior for points near the diagonal $i=j$ {is due to the fact that these} are affected by finite size effects {because of the different scaling behaviour of the cumulant on the diagonal}. [see Eq.~\eqref{g_i^4-c}]. }
\label{C_1^{2,1} and 2,2}
\end{figure}
 
 \section{Numerical verification of our theoretical predictions}
 We now test the validity of our operator method by numerically verifying the results that are obtained using the operator method and which are difficult to obtain via the direct method discussed in sec. \ref{direct-approach}. As mentioned while describing the model in sec.~\ref{sec:model}, a random fraction $\eta$ ( chosen from distribution $R(\eta)$) of the mass from one site gets transferred  to two neighbouring sites with equal probability at unit rate. The masses at the boundary sites are externally controlled to 
 maintain the (time independent) mass distribution at the left and right boundaries to be $P_L(g)$ and $P_R(g)$ respectively. We simulate this dynamics using the Monte-Carlo method. In our simulation we choose the following jump distribution 
 \be
 R(\eta) =\Theta(1-\eta) \Theta(\eta) \left[3~\Theta\left( \frac{1}{4}-\eta\right) + \frac{1}{3} \Theta \left(\eta-\frac{1}{4}\right)\right ]
 \label{R(eta)}
 \ee
where $\Theta(x)$ is the Heaviside  theta function. For this distribution we get
\beq
\mu_1=\frac{1}{4},~\mu_2=\frac{1}{8},~\mu_3=\frac{11}{128} \mbox{ and } \mu_4= \frac{43}{640}. \label{mus-simul}
\eeq
At the boundaries, we consider the following mass distributions
\be
P_{L,R}(G)= \frac{1}{g_{L,R}} e^{-G/g_{L,R}}
\ee
with $g_L = 2$ and $g_R = 1$. Below we provide numerical verification of our results for cumulants {of the first few orders}. We focus on two-site cumulants, which can be reduced to the functions $C_0^{(1)}(x)$ and $C_0^{(1,1)}(x,y)$ by the operator method. For the functions $C_0^{(1)}(x)$ and $C_0^{(1,1)}(x,y)$, we use the expressions in eqns. \eqref{av-m-ex-sol} and \eqref{C_ij-sol}.

\begin{figure}[h]
	\centering	
	\subfigure[]{
		\includegraphics[width=0.46\linewidth]{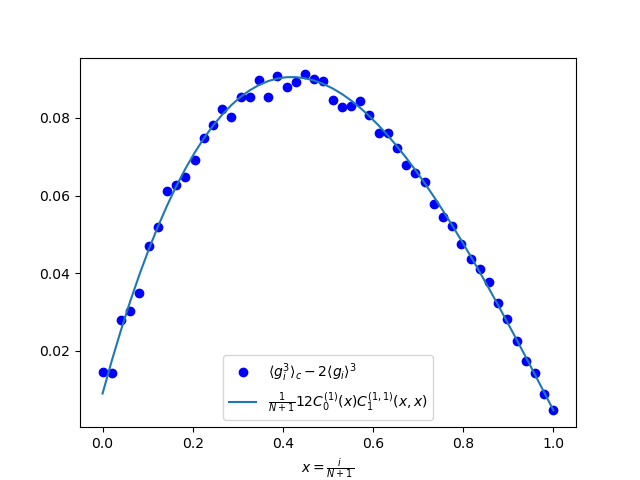}
	}
	\subfigure[]{
		\includegraphics[width=0.46\linewidth]{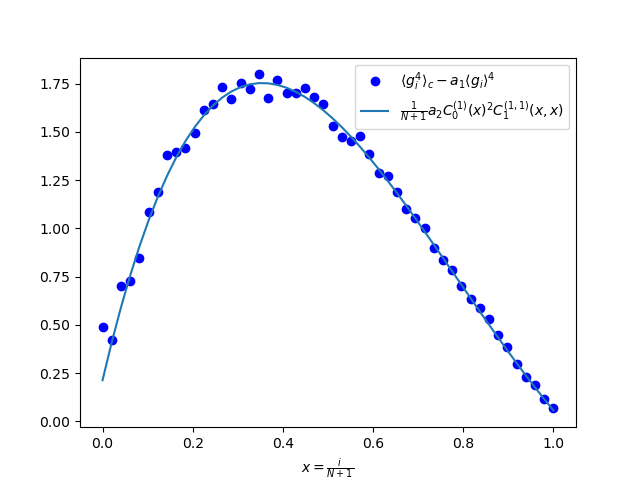}
	}
	\caption{Comparison with theory (solid line) of next-to-leading order ($O(\frac{1}{N})$) contributions to the one-site cumulants (a) $\la g_i^3 \ra$ and (b) $\la g_i^4 \ra$, with the $O(1)$ contributions subtracted. In (b), we use the numerical values $a_1 = \frac{4588}{499}$ and $a_2 = \frac{45492}{499}$  obtained from Eqs. \eqref{g_i^4-c} and \eqref{mus-simul}. The parameters used in the simulation are same as fig.~\ref{mean}.}
	\label{C3_C4}
\end{figure}

\begin{enumerate}
\item{} Mean $\langle g_i\rangle$: 
In fig.~\ref{mean}  we verify our first results on the mean mass $\la g_i \ra$. As announced in Eqs.~ \eqref{av-m-ex-sol} and \eqref{sol-C_0^(1)}, we indeed observe that the average mass decreases linearly as one goes from the left to the right end of the lattice. 
\item{} Correlation function $\la g_i g_j\ra_c$:
The scaling behaviour of  the two point connected correlation $\la g_i g_j \ra_c\vert_{i\neq j}$ as announced in Eqs.~\eqref{explicit-scaling-functions-C_ij} is verified in fig.~\ref{C_0^1-and-C_1^11} (a), where we plot the correlation along the line $i+j = N$.
\item{} {The variance $\la g_i^2 \ra_c$:
The behaviour of the variance as expected from Eq. \eqref{var-g_i} is shown in fig.~\ref{C_0^1-and-C_1^11} (b), where we have subtracted the $O(1)$ contribution $\frac{\mu_2}{\mu_1-\mu_2} C_0^{(1)}(x)^2$.}

\noindent These results which follow from the operator method were also obtained through the exact calculation in section III. We now move to higher cumulants, which cannot be easily derived from the microscopic method, but expressions for which were derived using the operator method in section IV.

\item{} The third-order cumulant on two sites, $\la g_i^2 g_j \ra_c$:
Next we consider the third-order cumulant $\la g_i^2 g_k \ra_c$  given in Eq.~\eqref{Cc-g_i-sq-g_k}. Note that this is the first non-trivial cumulant that we have calculated using the  operator method. In fig.~\ref{C_1^{2,1} and 2,2}(a),  we plot the leading order contribution to $\la g_i^2 g_k \ra_c$ given by the first term on the right hand side of Eq.~\eqref{Cc-g_i-sq-g_k}  along the line $i + j = N$. The excellent agreement between the theoretical prediction and simulation results validates our result.
\item{} The fourth-order cumulant on two sites, $\la g_i^2 g_j^2 \ra_c$:
In fig.~\ref{C_1^{2,1} and 2,2}(b) we plot the simulation results for cumulant $\la g_i^2 g_j^2 \ra_c$ along the line $i+j=N$ and compare it with the predictions from the operator method given in Eq.~\eqref{C_1^(22)}. We observe excellent agreement once again, except along the diagonal, $i=j$, where the scaling is different because the cumulant becomes $\la g_i^4 \ra_c$.  We examine this cumulant separately below.

\item{} The one-site third-order cumulant, $\la g_i^3 \ra_c$:
As given in Eq.~\eqref{g_i^3_c}, the $O(1)$ contribution to $\la g_i^3 \ra_c$ for the jump distribution in Eq.~\eqref{R(eta)} is equal to $2 C_0^{(1)}(x)^3$. Subtracting this contribution from $\la g_i^3 \ra_c$, we compare the $O(\frac{1}{N})$ correction with simulation results in fig.~\ref{C3_C4}(a). This verifies the validity of the operator method for higher-order cumulants.
\item{} The one-site fourth-order cumulant, $\la g_i^4 \ra_c$:
We finally compare the next-to-leading-order prediction ({\it{i.e}} the term at $O(\frac{1}{N})$) for the fourth cumulant $\la g_i^4 \ra_c$, as given in Eq.~\eqref{g_i^4-c} with simulation results in fig.~\ref{C3_C4}(b). Once  again the excellent agreement shows that the operator method captures higher cumulants very well.
\end{enumerate}

\section{Moment generating functions and derivation of the Poisson equations for the max-site correlations} \label{sec:Poisson}
\label{sec:cgf}
In this section we look at the the moment generating function (MGF) of the masses, defined as 
\bea
\left \la e^{\sum_{i=1}^N\lambda_i g_i} \right \ra &=& \sum_{m=0}^\infty \frac{1}{m!} \left \la \left( \lambda_1g_1 
+ \lambda_2 g_2 + ....+\lambda_Ng_N \right)^m \right \ra
\nonumber \\&=& 
\sum_{m=0}^\infty~
\underbrace{\sum_{m_1=0}^m\sum_{m_2=0}^m...\sum_{m_N=0}^m}_{\sum_{i=1}^Nm_i=m}\frac{1}{m_1!m_2!...m_N!}~\lambda_1^{m_1}\lambda_2^{m_2}...\lambda_N^{m_N} \la g_1^{m_1} g_2^{m_2}...g_N^{m_N}\ra.~~~~~~
\label{cfg-g_i-all}
\eea
For mass at single site, say $i$th, the MGF is 
\bea
\left \la e^{\lambda g_i} \right \ra &=& \sum_{n=0}^\infty \frac{1}{n!} \lambda^n \left \la g_i^n \right \ra,
\label{cfg-g_i}
\eea
which using the operator equation \eqref{op-split} can be written as 
\be
\left \la e^{\lambda g_i} \right \ra = \sum_{n=0}^\infty \frac{1}{n!} \lambda^n f_n\left \la \overbrace{g_i \cdot g_i\cdot...\cdot g_i}^{n~times} \right \ra,
\label{cfg-g_i-1}
\ee
Defining the function 
\be 
G(\lambda)= \sum_{n=0}^\infty \frac{f_n}{n!} \lambda^n \label{def-G-fn} 
\ee 
it is easy to see that the MGF in Eq.~\eqref{cfg-g_i-1} can be written as 
\bea
\left \la e^{\lambda g_i} \right \ra &=& \left \la G \left(\lambda ~g_i\cdot \right) \right \ra,~~\text{where} ~~
G \left(\lambda ~g_i\cdot \right) = \sum_{n=0}^\infty \frac{f_n}{n!} \lambda^n \overbrace{g_i \cdot g_i\cdot...\cdot g_i}^{n~times}, 
\label{G-op}
\eea
represents the MGF operator. We now take the extra step of assuming that the operator equation \eqref{op-split} is valid to all orders in $1/N$. This means that the above equation is assumed to be exact. This is warranted because, as we will see below, the terms in the steady-state equations for the CGF which remain after the operator recursion is satisfied can be incorporated into source terms for the Poisson equations ``max-site" correlation functions. In the following, we demonstrate this explicitly for the 2-point and 3-point correlation functions. In other words, we conjecture that in the continuum limit, the RATP steady-state is fully described by the operator recursion equations \eqref{op-split} and Poisson equations for the correlation functions, eqns. \eqref{Poisson-ell} and \eqref{BC-poisson-ell}.

For convenience, we now define a new operator $(\delta_i.)$, such that $g_i \cdot=m_i+\delta_i\cdot$ where $m_i=\la g_i \ra$, one gets 
\bea
\left \la e^{\lambda g_i} \right \ra &=& \left \la G \left(\lambda (m_i+\delta_i\cdot)\right) \right \ra,~~\text{and~for~the~joint~CGF} \label{S-CGF-op} \\
\left \la e^{\sum_{i=1}^N\lambda g_i} \right \ra &=& \left \la G \left(\lambda_1 (m_1+\delta_1\cdot \right )G \left(\lambda_2 (m_2+\delta_2\cdot \right ) ... G \left(\lambda_N (m_N+\delta_N \cdot \right )\right \ra. \label{J-CGF-op} 
\eea
The advantage of separating the mean $m_i$ from $g_i$ is that the average of the combinations of $\delta_i$'s  directly represents the cumulants  and hence the scaling correlation functions in the large $N$ limit e.g. 
\bea
\la \delta_i\ra &=& \la g_i \ra_c = 0, \label{av-D_i} \\
\la \delta_i\delta_j\ra_{i\neq j} &=&  \la g_i g_j\ra_c \big{|}_{i\neq j} \to \frac{1}{N} C_1^{(1,1)}(x,y),~~\text{with}~~x=\frac{i}{N},~y=\frac{j}{N}, \label{av_D_i-D_j} \\
\la \delta_i\delta_j\delta_k\ra_{i\neq j \neq k} &=&  \la g_i g_jg_k\ra_c\big{|}_{i\neq j \neq k}  \to \frac{1}{N^2} C_2^{(1,1,1)}(x,y,z), 
~~\text{with}~~x=\frac{i}{N},~y=\frac{j}{N}~\text{and}~z=\frac{k}{N}.
 \label{av_D_i-D_j-D_k} 
\eea
If the CGF's were known explicitly then one can expand both sides of Eq.~\eqref{S-CGF-op} and \eqref{J-CGF-op} in powers of $\lambda$s. For example, the right hand side(rhs) of one-point CGF can be expanded as 
\bea
 \left \la G \left(\lambda (m_i+\delta_i\cdot)\right) \right \ra = G(\lambda m_i) + \frac{1}{2} \lambda^2 G''(\lambda m_i) \la \delta_i \cdot \delta_i \ra + ...,
\eea
whereas the expansion of the left hand side(lhs) is given in Eq.~\eqref{cfg-g_i}. In the above we have used  $G''(z)=d^2 G(z)/dz^2$ and $\la \delta_i \ra=0$.
For two-point CGF also, the rhs can be expanded similarly and one has 
\bea
 \left \la G \left(\lambda_i (m_i+\delta_i\cdot)\right) G \left(\lambda_j (m_j+\delta_j\cdot)\right)\right \ra_{i \neq j}  = \left \la G \left(\lambda_i (m_i+\delta_i\cdot)\right) \ra \la G \left(\lambda_j (m_j+\delta_j\cdot)\right)\right \ra + \lambda_i \lambda_j G'(\lambda_im_i)G'(\lambda_jm_j)~\la \delta_i \cdot \delta_j \ra +... ~~~~~~
\eea
Similar expressions hold for multi-site cumulants. 
Equating terms of same order from both sides, one can in principle obtain the connected cumulants like $\la g_i g_j\ra_c$, $\la g_i^m g_j^n\ra_c$ etc. at different orders of $1/N$, in terms of the corresponding scaling correlation functions $C_{\ell-1}^{(1,1,1,...)}(x,y,z,...)$. It turns out that it is difficult to compute the MGFs exactly however  it is possible to compute them order by order in $\lambda$s and such an attempt, in each order provide the Poisson equation (mentioned in Eq.~\eqref{Poisson-ell}) satisfied by the max-site correlation at that order. In the following we demonstrate how does one get such Poisson equation at order $\ell=2$ and $\ell=3$.

\subsection{Derivation of the Poisson equation satisfied by $C_1^{(1,1)}(x,y)$}

\subsubsection{MGF equation of  $g_i$ }
\label{2pt-diag-eqns}
To proceed, starting from the master equation \eqref{eq:Pt} we obtain the following equation satisfied by the MGF $\la e^{\lambda g_i}\ra$ in the steady state:
\bea
4 \la e^{\lambda g_i}\ra = \left \la e^{\lambda (g_i+\eta g_{i+1})} \right \ra+ \left \la e^{\lambda (g_i+\eta g_{i-1})} \right \ra +2  \left\la e^{\lambda (1-\eta)g_{i}} \right \ra
\label{1-pt-mgf-eq}
\eea
In the above equation,  $\la ... \ra$ on terms involving $\eta$ also represents average over $\eta$ in addition to that over the distribution of $g_i$ \emph{i.e.} $ \left\la e^{\lambda (1-\eta)g_{i}}\right \ra \equiv \int_0^1 d\eta R(\eta)~ \left\la e^{\lambda (1-\eta)g_{i}} \right \ra$.
Note that this equation can be rewritten in terms of the $G$ function, introduced in Eq.~\eqref{def-G-fn}, as 
\be
4 \la G(\lx g_i\cdot)\ra -2 \la G(\lx (1-\eta) g_i\cdot) \ra = \la G(\lx g_{i}\cdot)G(\lx \eta g_{i+1}\cdot)\ra +\la G(\lx g_{i}\cdot)G(\lx \eta g_{i-1}\cdot)\ra
\label{G-S-CGF-eq}
\ee

Now, we show that the recursion relations\eqref{fn-recur} get translated to the operator equation 
\be
2 \la G(\lx g_i\cdot)\ra - \la G(\lx (1-\eta) g_i\cdot) \ra = \la G(\lx g_{i}\cdot)G(\lx \eta g_{i} \cdot)\ra 
\label{av-G-rel}
\ee
This can be seen by using Eq.~ \eqref{cfg-g_i-1} to get
\beqa
2 \sum_n \frac{1}{n!} \lx^n f_n - \sum_n \frac{1}{n!} \lx^n f_n \la (1-\eta)^n \ra = \sum_i \sum_j \frac{1}{i!} \frac{1}{j!} \lx^{i+j} \la \eta^j \ra f_i f_j
\eeqa
Equating coefficients of $\lambda^n$ on both sides, we get
\beq
f_n (2 -\la (1-\eta^n) \ra ) = f_0 f_n + \sum_{k=1}^{n-1} \binom{n}{k} \mu_k f_k f_{n-k} + \mu_n f_0 f_n 
\eeq
which, recalling that $f_0=1$, is equivalent to Eq.~ \eqref{fn-recur}.

We now demand that the operator equation Eq.~ \eqref{G-S-CGF-eq} is valid in all orders of $\lx$, that is,
\be
\la \Delta G \ra =0,~\text{where},~
\Delta G=G(\lx g_i\cdot) G(\lx \eta g_{i+1}) + G(\lx g_i\cdot) G(\lx \eta g_{i-1} \cdot) -2 G(\lx g_i\cdot) G(\lx \eta g_{i} \cdot) 
\label{DeltaG=0}
\ee
for large $N$. Using the  expansion of $G(\lx g_i.)$ given in Eq.~\eqref{def-G-fn} in equation \eqref{DeltaG=0} and then equating terms of different orders of $\lx$ on the lhs to zero, we get relations among various cumulants as observed in sec.~\ref{sec:operator}.

\begin{itemize}
\item[--] $O(\lx^0)$:  At this order we get the identity $2=2$.
\item[--] $O(\lx)$: At this order we get 
\be
\la g_{i+1}\ra -2 \la g_i \ra +\la g_{i-1}\ra=0
\ee
which simplifies to 
\beq
 \nabla_i^2 m_i \implies \partial_x^2 C_0^{(1)}(x) = 0,
 ~~\text{where}~~x=\frac{i}{N}~~\text{with}~~C_0^{(1)}(0)=g_L,~\text{and}~C_0^{(1)}(1)=g_R.
\eeq
Here $ \nabla_i^2$ denotes the discrete second difference operator \emph{i.e.} 
$\nabla_i^2 f_i = f_{i+1} -2 f_i +f_{i-1}$,
and $\partial_x^2 \equiv N^2\Delta_i^2$ represent the continuum partial derivative of second order.
It is easy to check from Eq.~\eqref{sol-C_0^(1)} that the above equation is indeed true. 

\item[--] $O(\lx^2)$: Collecting terms at $O(\lx^2)$ from both sides and equating we get
\begin{align}
\begin{split}
 \mu_1f_1\left[ \la g_i \cdot g_{i+1}\ra +\la g_{i+1}  \cdot g_i \ra+  \la g_i \cdot g_{i-1}\ra \right. &+ \left. \la g_{i-1} \cdot g_{i}\ra- 4 \la g_i\cdot g_i \ra \right] \\ 
&+ \mu_2 f_2 \left[ \la g_{i+1}\cdot g_{i+1} \ra + \la g_{i-1} \cdot g_{i-1}\ra - 2 \la g_{i} \cdot g_i \ra \right] = 0,
 \end{split}
 \label{diag-eq}
\end{align}
which, after using $g_i=m_i+\delta_i$ simplifies to 
\be
\left[ \la \delta_i \cdot \delta_{i+1} \ra + \la \delta_{i+1} \cdot \delta_{i} \ra + \la \delta_i \cdot \delta_{i-1} \ra + \la \delta_{i-1} \cdot \delta_{i} \ra - 4 \la \delta_i \cdot \delta_i \ra\right] = -\frac{\mu_2}{(\mu_1-\mu_2) } \left[  \nabla_i^2 m_i^2 + \nabla_i^2 \la \delta_i \cdot \delta_i \ra \right],
\label{diag-C_ii}
\ee
where $\nabla_i^2$ is defined above.
\end{itemize}

\subsubsection{Joint MGF equations of $g_i$ and $g_{i \pm1}$ }
\label{2pt-above-diag-eqns}
Till now we have looked at how the correlations $\la \delta_i \cdot \delta_j \ra$ near the diagonal $i=j$ behave. Let us now look at the equation satisfied by the joint MGF of $g_i$ and $g_{i+1}$, which can again be obtained from the master equation \eqref{eq:Pt}:
\begin{align}
6\left \la e^{\lx_1g_i+\lx_2g_{i+1}}\right\ra = &\left \la e^{\lx_1g_i+[\lx_2+\eta (\lx_1-\lx_2)]g_{i+1}}\right\ra  
+ \left \la e^{[\lx_1-\eta (\lx_1-\lx_2)]g_{i}+\lx_2g_{i+!}}\right\ra + \left \la e^{\lx_1g_i+\lx_1\eta g_{i-1}+\lx_2g_{i+1}}\right\ra \nonumber \\
&+ \left \la e^{\lx_1g_i+\lx_2g_{i+1}+\lx_2\eta g_{i+2}}\right\ra + \left \la e^{\lx_1(1-\eta)g_i+\lx_2g_{i+1}}\right\ra
+ \left \la e^{\lx_1g_{i}+\lx_2(1-\eta)g_{i+1}}\right\ra.
\label{2-pt-mgf-i_i+1-eq}
\end{align}
Expressing this equation in terms of the $G$ function, introduced in Eq.~\eqref{def-G-fn}, and simplifying using Eqs.~\eqref{av-G-rel} and \eqref{DeltaG=0}, we get 
\begin{align}
\begin{split}
8 \la G(\lx_1g_i\cdot)G(\lx_2g_{i+1}\cdot) \ra 
&=  \la G(\lx_1g_i\cdot)G(\lx_1\eta g_{i+1}\cdot)G(\lx_2g_{i+1}\cdot) \ra + \la G(\lx_1g_i\cdot)G(\lx_1\eta g_{i-1}\cdot)G(\lx_2g_{i+1}\cdot) \ra  \\ 
&+\la G(\lx_1g_i\cdot)G(\lx_2g_{i+1}\cdot) G(\lx_2\eta g_{i+2}\cdot)\ra 
+\la G(\lx_1g_i\cdot)G(\lx_2g_{i+1}\cdot) G(\lx_2\eta g_{i}\cdot)\ra  \\
&+2 \la G(\lx_1[1-\eta]g_i\cdot)G(\lx_2g_{i+1}\cdot)\ra
+2 +\la G(\lx_1g_i\cdot)G(\lx_2[1-\eta]g_{i+1}\cdot) \ra.
\end{split}
\end{align}
Following a similar procedure as done above, one finds that equations at $O(\lx_1)$, $O(\lx_2)$, $O(\lx_1^2)$ and $O(\lx_2^2)$ are already obtained in the sec. \ref{2pt-diag-eqns}. We obtain a new equation only by equating the coefficient of  $O(\lx_1\lx_2)$ term to zero.
$\la g_i\cdot g_{i+2}\ra +\la g_i\cdot g_{i}\ra +\la g_{i+1}\cdot g_{i+1}\ra +\la g_{i-1}\cdot g_{i+1}\ra-4 +\la g_{i}\cdot g_{i+1}\ra =0$
which, after using the form $g_i=m_i+\delta_i$ simplifies to 
\be
\la \delta_i\cdot \delta_{i+2}\ra +\la \delta_i\cdot \delta_{i}\ra +\la \delta_{i+1}\cdot \delta_{i+1}\ra +\la \delta_{i-1}\cdot \delta_{i+1}\ra-4 \la \delta_{i}\cdot \delta_{i+1}\ra =0.
\label{eq_C_i_i+1}
\ee

\subsubsection{Joint MGF equations of $g_i$ and $g_j$ with $|i - j| \geq 2$}
\label{2pt-bulk-eqns}
To see how the correlations $\la \delta_i \delta_j \ra$  in the bulk \emph{i.e.} $i \neq j$ behave, we consider from the following two-point MGF equation
\bea
\begin{split}
8 \left \la e^{\lx_1g_i+\lx_2g_j}\right \ra &=& \left \la e^{\lx_1(g_i+\eta g_{i+1})+\lx_2g_j}\right \ra + \left \la e^{\lx_1(g_i+\eta g_{i-1})+\lx_2g_j}\right \ra 
+\left \la e^{\lx_1g_i+\lx_2(g_j+\eta g_{j+1})}\right \ra + \left \la e^{\lx_1g_i+\lx_2(g_j+\eta g_{j-1})}\right \ra \\
&& +2 \left \la e^{\lx_1(g_i-\eta g_i)+\lx_2g_j}\right \ra + 2 \left \la e^{\lx_1g_i + \lx_2(g_j-\eta g_j)}\right \ra.
\end{split}
\label{2-pt-mgf-eq}
\eea
In this case also we do not get any new equations from $O(\lx_1)$, $O(\lx_2)$, $O(\lx_1^2)$ and $O(\lx_2^2)$ as expected. The only new equation we get from  $O(\lx_1\lx_2)$ is 
$\la g_i\cdot g_{j+1}\ra +\la g_i\cdot g_{j-1}\ra +\la g_{i+1}\cdot g_j\ra +\la g_{i-1}\cdot g_j\ra-4 +\la g_{i}\cdot g_j\ra =0$
which, after using the form $g_i=m_i+\delta_i$ simplifies to 
\be
\la \delta_i\cdot \delta_{j+1}\ra +\la \delta_i\cdot \delta_{j-1}\ra +\la \delta_{i+1}\cdot \delta_j\ra +\la \delta_{i-1}\cdot \delta_j\ra-4 \la \delta_{i}\cdot \delta_j\ra =0.
\label{bulk_C_ij}
\ee

Combining the  Eqs.~\eqref{diag-C_ii}, \eqref{eq_C_i_i+1} and \eqref{bulk_C_ij}, appropriately, we get
\begin{align}
\mathcal{D}_{ij} \la \delta_i \cdot \delta_j \ra
&= -\delta_{ij}~\frac{\mu_2}{\mu_1-\mu_2}~(\nabla^f_i m_i)^2~\left[  \nabla_i^2 m_i^2 + \nabla_i^2 \la \delta_i \cdot \delta_i \ra \right],~~\text{where},
\label{corr-eq-2d} \\
\mathcal{D}_{ij} \la \delta_i \cdot \delta_j \ra&=\la \delta_i\cdot \delta_{j+1}\ra +\la \delta_i\cdot \delta_{j-1}\ra +\la \delta_{i+1}\cdot \delta_j\ra 
+\la \delta_{i-1}\cdot \delta_j\ra-4 \la \delta_{i}\cdot \delta_j\ra 
\end{align}
and $\delta_{ij}$ is Kronecker delta. We have used the notation $\nabla^f_i$ to denote the forward difference operator, $\nabla_i^f f_i \equiv f_{i+1} - f_i$. Our next task is to go to the the continuum limit in the $N \to \infty$ limit where we replace
\beqa
\begin{split}
\delta_{ij} &\rightarrow& \frac{1}{N}\delta(x-y) \\
(\nabla^f_i m_i)^2 &\rightarrow& \frac{2}{N^2} (\partial_x C_0^{(1)}(x))^2 \\
\mathcal{D}_{ij} \la \delta_i \cdot \delta_j \ra &\rightarrow& \frac{1}{N^3} 
\nabla^2 C_1^{(1,1)}(x,y)
\end{split}
\label{continum-limit}
\eeqa
where $\nabla^2=\partial_x^2+\partial_y^2$ denotes the continuum Laplacian. 
Inserting these continuum forms in Eq.~\eqref{corr-eq-2d} we in the leading order $O(1/N^3)$ get, 
\beq
\nabla^2 C_1^{(1,1)}(x,y) = -\delta(x-y)~\frac{2 \mu_2}{\mu_1-\mu_2} (\partial_x C_0^{(1)}(x))^2 .
\eeq
One needs to solve this equation inside a the unit square $(x,y) \in[0,1] \times [0,1]$ with the correlation function vanishing at the boundaries \emph{i.e.} $C_1^{(1,1)}(x,y)\vert_{(x,y) \in \partial} =0$. This is because the fluctuations of the masses at the boundaries 
are independent of the bulk masses. The solution to this equation is easily found to be
	\beqa
	C_1^{(1,1)}(x,y) = 
	\begin{cases}
		\frac{ \mu_2}{\mu_1-\mu_2} (g_R - g_L)^2 y(1-x) &\mbox{ if } x>y \\
		\frac{ \mu_2}{\mu_1-\mu_2} (g_R - g_L)^2 x(1-y) &\mbox{ if } y>x
	\end{cases}
	\eeqa
	as already announced in Eq.~\eqref{explicit-scaling-functions-C_ij}.

\subsection{Derivation of the Poisson equation satisfied by $C_2^{(1,1,1)}(x,y,z)$}

We now outline the derivation of the Poisson equation satisfied by the three-point scaled correlation function. We first consider the case when two of the points are coincident, and using the equation for the cumulant $\la g_k e^{\lambda g_i} \ra$ at $O(\lambda^2$), with points $i$ and $k$ far away from each other. Then we consider when all three points are coincident, and thus consider $\la e^{\lambda g_i} \ra$ at $O(\lambda^3)$.

\subsubsection{Two coincident points, $y=z$}

Consider the evolution equation for $\la g_k e^{\lambda g_i} \ra$, where $|i-k|\ge 2$,
\beq
\frac{d}{dt}\la g_k e^{\lambda g_i} \ra = \la (\nabla^f_k g_k + \nabla^b_k g_k) e^{\lambda g_i} \ra + \la g_k (\Delta G)_i \ra
\eeq
where $\Delta G$ is the operator defined previously in Eq.~ \eqref{DeltaG=0}, and the subscript $i$ denotes that it is taken at the site $i$. We also use the notation $\nabla^f_i$ and $\nabla^b_i$ for forward and backward difference operators, $\nabla^f_i f_i \equiv f_{i+1} - f_i$, $\nabla^b_i f_i \equiv f_{i-1} - f_{i}$. Setting the LHS to $0$ we get the steady-state values. Further, using $g_k = m_k + \delta_k$, with $\nabla_k^2 m_k = 0$ and $m_k \la (\Delta G)_i \ra = 0$, we get 
\beq
\la \eta\nabla_k^2 \delta_k e^{\lambda g_i} \ra + \la \delta_k (\Delta G)_i \ra = 0
\eeq
We can now expand in orders of $\lambda$ to get equations for the correlation functions. At $O(\lambda)$ we get $\mathcal{D}_{ki} \la \delta_k \delta_i \ra$, as previously attested. At $O(\lambda^2)$, the non-zero contribution from the first term is
\beq
\mu_1 \frac{f_2}{2} \la (\nabla_k^2  \delta_k) (2 m_i \delta_i + \delta_i^2)\ra
\eeq
and that from the second term is 
\beq
\bigg\la \delta_k \eta [ f_1^2 + f_2 \eta ] g_i [\nabla_i^f g_i + \nabla_i^b g_i]  + \frac{f_2}{2} \eta^2 [(\nabla_i^f g_i)^2 + (\nabla_i^b g_i)^2] \bigg\ra
\eeq
After taking the continuum limit for large $N$, as done in Eq.~\eqref{continum-limit}, we get,
\beq
\nabla^2 C(x,y,z)\vert_{y=z} = -\frac{2 \mu_1}{3 \mu_1-2 \mu_2} (\partial_y M(y)) (\partial_y C(x,y))
\eeq

\subsubsection{Three coincident points, $x=y=z$}
For all three points coincident, the derivation involves keeping terms in $\la \Delta G \ra = 0$ to order $\lambda^3$. At $O(\lambda^3)$, the terms in $\Delta G$ are
\beqa
&\bigg[&(\frac{\mu_1}{2} + \mu_2) f_2 + \frac{1}{3} f_3 \mu_3 \bigg] \bigg[ 2 m_i \la \delta_i \nabla_i^2  \delta_i \ra + \la \delta_i^2 \nabla_i^2 \delta_i \ra\bigg]\nonumber\\
+ &\bigg[& \frac{1}{2} \mu_2 f_2 + \frac{1}{2} \mu_3 f_3 \bigg] \bigg[ 2 m_i (\nabla_i^f m_i)^2 + \la (m_i + \delta_i) ((\nabla_i^f (m_i + \delta_i))^2 + (\nabla_i^b (m_i + \delta_i))^2) \ra\bigg] \nonumber\\
+ &\bigg[& \frac{1}{6} \mu_3 f_3 \bigg] \bigg[ \la ((\nabla_i^f (m_i + \delta_i))^2 + (\nabla_i^b (m_i + \delta_i))^2 \ra\bigg].
\eeqa
In the following we use the below expressions in the continuum limit,
\beqa
\la \delta_i \nabla_i^2 \delta_i \ra &\rightarrow& \frac{1}{N^3}\frac{1}{2} \nabla^2 C(x,y) \vert_{x=y}\\
\la (\nabla_i^f \delta_i)^2 +  (\nabla_i^b \delta_i)^2 \ra &\rightarrow& - \frac{1}{N^3} \nabla^2 C(x,y) \vert_{x=y}\\
\la \delta_i^2 \nabla_i^2 \delta_i \ra &\rightarrow&\frac{1}{N^4} \frac{1}{3} \nabla^2 C(x,y,z) \vert_{x=y=z}\\
\la \delta_i ((\nabla_i^f \delta_i)^2 +  (\nabla_i^b \delta_i)^2) \ra &\rightarrow& - \frac{1}{N^4} \nabla^2 C(x,y,z) \vert_{x=y=z}\\
\la (\nabla_i^f \delta_i)^3 + (\nabla_i^b \delta_i)^3 \ra &\rightarrow& 0 + O(\frac{1}{N^5}).
\eeqa
We also use that $f_1 = 1$, $f_2 = \frac{\mu_1}{\mu_2-\mu_1}$ and $f_3 = \frac{\mu_1 + \mu_2}{\mu_1-\mu_2} f_2$. Simplifying, we get 

\beq
\nabla^2 C(x,y,z)\vert_{x=y=z} = k_3 \partial_x^2 (M(x)^3)
\eeq
where
\beq
k_3 = \frac{\mu_2}{\mu_1-\mu_2} \frac{2 \mu_2^2 - \mu_3 (\mu_1 + \mu_2)}{\mu_1(\mu_1-\mu_2) - \mu_3 (\mu_1+\mu_2)}.
\eeq
Thus, collecting all the possible source terms, we finally get the following  equation for the 3-point correlation function,
\beqa
\nabla^2 C(x,y,z) &=& -\frac{2 \mu_1}{3 \mu_1-2 \mu_2} \big[ (\partial_y M(y)) (\partial_y C(x,y)) \delta(y-z) + (\partial_x M(x)) (\partial_x C(x,z)) \delta(x-y) + (\partial_z M(z)) (\partial_z C(y,z)) \delta(x-z)\big] \nonumber\\
 & & + \frac{\mu_2}{\mu_1-\mu_2} \frac{2 \mu_2^2 - \mu_3 (\mu_1 + \mu_2)}{\mu_1(\mu_1-\mu_2) - \mu_3 (\mu_1+\mu_2)} \partial_x^2 (M(x)^3) \delta(x-y) \delta(y-z)
\eeqa

\section{Conclusion}
In this paper we have studied a mass transfer model defined on a one dimensional lattice of size $N$ driven by two reservoirs of different `chemical potentials' at the two ends. We have shown that in the steady state, correlations involving a arbitrary set of sites (say $k$ sites out of $N$) and arbitrary order, say $M$, {have}	 expansions in powers of $1/N$ starting at order $O(1/N^{k-1})$ to order $O(1/N^{M-1})$. We found that at each order, the correlation functions possess scaling forms which can in principle be expressed explicitly in terms of lower order max-site correlations. This is in contrast to Wick's theorem where any higher order correlations can be expressed in terms of two point correlation functions.  The important difference is that the scaling function of a $k-$point correlation of arbitrary order involves all the max-site correlations $\la g_{i_{k_1}}g_{i_{k_2}}...g_{i_{k_m}}\ra_c$ upto order $k$ \emph{i.e.} with $1\leq m \leq k$. To establish this result we employed a novel operator method, valid in the large $N$ limit. The intriguing recursive structure of {the} operator method allows us to determine the scaling functions at different order explicitly {if} the max-site correlation functions up to  that order are known. These max-site correlation functions satisfy Poisson equations inside a hypercube. The operator method {also} provides a simple way to derive these Poisson equations for the two point and three point correlations. The operator method can in principle be employed to derive the explicit form of Poisson equation for higher point correlations systematically. 

Our study of multi-site correlations of masses at arbitrary order in the NESS can be extended in different directions. One immediate question {is} whether this special structure suggests exact solvability of this model. Another interesting question is whether a similar structure exists for correlations in non-stationary regimes as well. We also expect a similar operator method to hold in higher dimensions, and it would then be interesting to extend the results of this paper {to two- and three-dimensional versions of the RATP.} Finally, another interesting direction would be to explore the possibility of the existence of such an operator method in other mass transfer models.

\section{Acknowledgements}
The authors would like to acknowledge helpful discussions with Deepak Dhar, Kabir Ramola, Satya N. Majumdar and Arghya Das. R.D. acknowledges the hospitality received from ICTS during his visit  supported by the `Infosys Excellence Grant'.  This work was started during R.D.'s visit to ICTS. A.K. acknowledges the support of the core research grant no. CRG/2021/002455  and  MATRICS grant MTR/2021/000350  from the Science and Engineering Research Board (SERB), Department of Science and Technology, Government of India. A.K. also acknowledges support from the Department of Atomic Energy, Government of India, under project no.  19P1112R\&D.

	\appendix
	\section{Hierarchy of the correlations $\langle g_{i_1}^{m_1}g_{i_2}^{m_2}...g_{i_k}^{m_k}\rangle_c$ till order $M=3$}
	\label{hierarchy}
	{In the following, we use the notations $x \equiv \frac{i}{N+1}$, $y \equiv \frac{j}{N+1}$, $z \equiv \frac{k}{N+1}$ and $w \equiv \frac{l}{N+1}$.}
	
	\begin{itemize}
		\item Correlations of order $M=1$:
		\begin{align}
		\langle g_i \rangle \approx  C_0^{(1)}(x) \nonumber
		\end{align}
		\item Correlations of order $M=2$:
		\begin{align}
		\langle g_ig_j \rangle &\approx \frac{1}{N}~C_1^{(1,1)}(x,y) \nonumber \\
		\langle g_i^2 \rangle &\approx C_0^{(2)}(x) + \frac{1}{N}~C_1^{(2)}(x)  \nonumber 
		\end{align}
		\item Correlations of order $M=3$:
		\begin{align}
		\langle g_ig_j g_k\rangle &\approx \frac{1}{N^2}~C_2^{(1,1,1)}(x,y,z) \nonumber \\
		\langle g_i^2g_k \rangle &\approx \frac{1}{N}~C_1^{(2,1)}(x,y) +  \frac{1}{N^2}~C_2^{(2,1)}(x,y) \nonumber \\
		\langle g_i^3 \rangle &\approx C_0^{(3)}(x)+\frac{1}{N}~C_1^{(3)}(x) +  \frac{1}{N^2}~C_2^{(3)}(x) \nonumber 
		\end{align}
		\item Correlations of order $M=4$:
		\begin{align}
		\langle g_ig_j g_kg_l\rangle &\approx \frac{1}{N^3}~C_3^{(1,1,1,1)}(x,y,z,w) \nonumber \\
		\langle g_i^2g_jg_k \rangle &\approx \frac{1}{N^2}~C_2^{(2,1,1)}(x,y,z) +  \frac{1}{N^3}~C_3^{(2,1,1)}(x,y,z) \nonumber \\
		\langle g_i^2g_j^2 \rangle &\approx \frac{1}{N}~C_1^{(2,2)}(x,y)+ \frac{1}{N^2}~C_2^{(2,2)}(x,y) +  \frac{1}{N^3}~C_3^{(2,2)}(x,y) \nonumber \\
		\langle g_i^3g_j \rangle &\approx \frac{1}{N}~C_1^{(3,1)}(x,y) +\frac{1}{N^2}~C_2^{(3,1)}(x,y) +\frac{1}{N^3}~C_3^{(3,1)}(x,y)\nonumber \\
		\langle g_i^4 \rangle &\approx C_0^{(4)}(x)+\frac{1}{N}~C_1^{(4)}(x) +\frac{1}{N^2}~C_2^{(4)}(x) +\frac{1}{N^3}~C_3^{(4)}(x)\nonumber 
		\end{align}
	\end{itemize}
	
	\section{Recursion relations for the operator $g_i^n$}
	\label{op-gen-n}
	
	Starting from the master equation in \eqref{eq:Pt}, one finds 
	\bea
	\frac{d}{dt} \la g^n_i \ra = \int_0^1d\eta~R(\eta)~\left[~\la (g_i +\eta g_{i+1})^n\ra + \la (g_i +\eta g_{i-1})^n\ra 
	+2 (1-\eta)^n\la g_i^n \ra - 4 \la g_i^n\ra~\right]
	\eea
	
	Thus, in the steady-state, expanding the brackets we have 
	\bea
	2 \la g_i^n \ra = \sum_{k=1}^{n} \binom{n}{k} \mu_k \left[ \la g_i^{n-k} (g_{i+1}^k + g_{i-1}^k) \ra \right]+ 2 \int_0^1d\eta~R(\eta)(1-\eta)^n \la g_i^n \ra 
	\eea
	which simplifies to 
	\bea
	\la(1-(1-\eta)^n)\ra \la g_i^n \ra- \frac{1}{2} \left( \la g_{i-1}^n \ra + \la g_{i+1}^n \ra \right)= \sum_{k=1}^{n-1} \binom{n}{k} \mu_k \left[ \left \la g_i^{n-k} \frac{(g_{i+1}^k + g_{i-1}^k)}{2} \right \ra \right] 
	\eea
Taking the {scaling} limit (\emph{i.e.} $N \to \infty$ limit) implies  $\frac{1}{2} (g_{i-1}^k + g_{i+1}^k) \rightarrow g_i^k + O(N^{-2})$., which finally provides the operator equation
	\bea
	g_i^{n}(x) = \frac{1}{\la 1-(1-\eta)^k\ra - \mu_n}\sum_{k=1}^{n-1} \binom{n}{k} \mu_k g_i^{n-k} \cdot g_i^k
	\eea
	
	It can be checked that for the cases $n=2$, $n=3$ and $n=4$, this gives eqns. (\ref{var-g_i}), (\ref{g_i^3_c}) and (\ref{g_i^4-c}) respectively.
	
	\bibliographystyle{unsrt}
	\bibliography{rap}

\end{document}